\documentclass[twocolumn]{aastex631}

\def\CII  {[C{\small{II}}]}
\def\micron {$\mu$m}

\shorttitle{Molecular gas in the Sombrero galaxy}
\shortauthors{Sutter \& Fadda}
\graphicspath{{./}{Figures/}}
\begin{document}

\title{A Molecular Gas Ring Hidden in the Sombrero Galaxy}

\correspondingauthor{Jessica Sutter (moving to UCSD)}
\email{jessica.sutter93@gmail.com}

\author[0000-0002-9183-8102]{Jessica Sutter}
\affiliation{SOFIA Science Center, USRA, NASA Ames Research Center, M.S. N232-12 Moffett Field, CA 94035, USA}
\author[0000-0002-3698-7076]{Dario Fadda}
\affiliation{SOFIA Science Center, USRA, NASA Ames Research Center, M.S. N232-12 Moffett Field, CA 94035, USA}



\begin{abstract}
We present Herschel, ALMA, and MUSE observations of the molecular ring of Messier 104, also known as the Sombrero galaxy. These previously unpublished archival data shed new light on the content of the interstellar medium of M104. In particular, molecular hydrogen measured by CO emission and dust measured by far-infrared light are uniformly distributed along the ring. The ionized gas revealed by H$\alpha$ and \CII{} emission is distributed in knots along the ring. Despite being classified as an SAa galaxy, M104 displays features typical of early-type galaxies. We therefore compared its \CII{} and dust emission to a sample of early-type galaxies observed with {\it Herschel} and SOFIA. The \CII/FIR ratio of M104 is much lower than that of typical star-forming galaxies and is instead much more similar to that of early-type galaxies. By classifying regions using optical emission line diagnostics we also find that regions classified as HII lie closer to star-forming galaxies in the \CII/FIR diagram than those classified as low-ionization emission regions. The good match between \CII{} and H$\alpha$ emission in conjunction with the lack of correlation between CO emission and star formation suggest that there is very limited active star formation along the ring and that most of the \CII{} emission is from ionized and neutral atomic gas rather than molecular gas. From the total intensity of the CO line we estimate a molecular hydrogen mass of $0.9\times10^{9}$M$_\odot$, a value intermediate between those of early type galaxies and the content of the molecular ring of our galaxy.

\end{abstract}



\section{Introduction} \label{sec:intro}
Galaxies are complex structures that can be shaped by a variety of factors.  Each galaxy has its own unique formation history, which can lead to the creation of spiral arms, rings, bars, or even a lack of structure all together.  In addition to the distinct structural nature of each galaxy, our observing perspective can have significant effects on the measurable properties of a galaxy.  A dusty, edge-on disk galaxy may appear devoid of optical tracers of star-formation from our viewing angle, but could present a different picture if viewed face-on. Piecing together observations of a single galaxy across the electromagnetic spectrum provides the context to disentangle different structures and determine how viewing angles affect our ability to distinguish galaxy characteristics.

%Variations in composition can play important roles in determining how efficiently a galaxy is forming stars, how long star formation will last, and even effect the validity of certain interstellar medium diagnostics.

\begin{figure*}[t!]
\begin{center}
\includegraphics[width=0.9\textwidth]{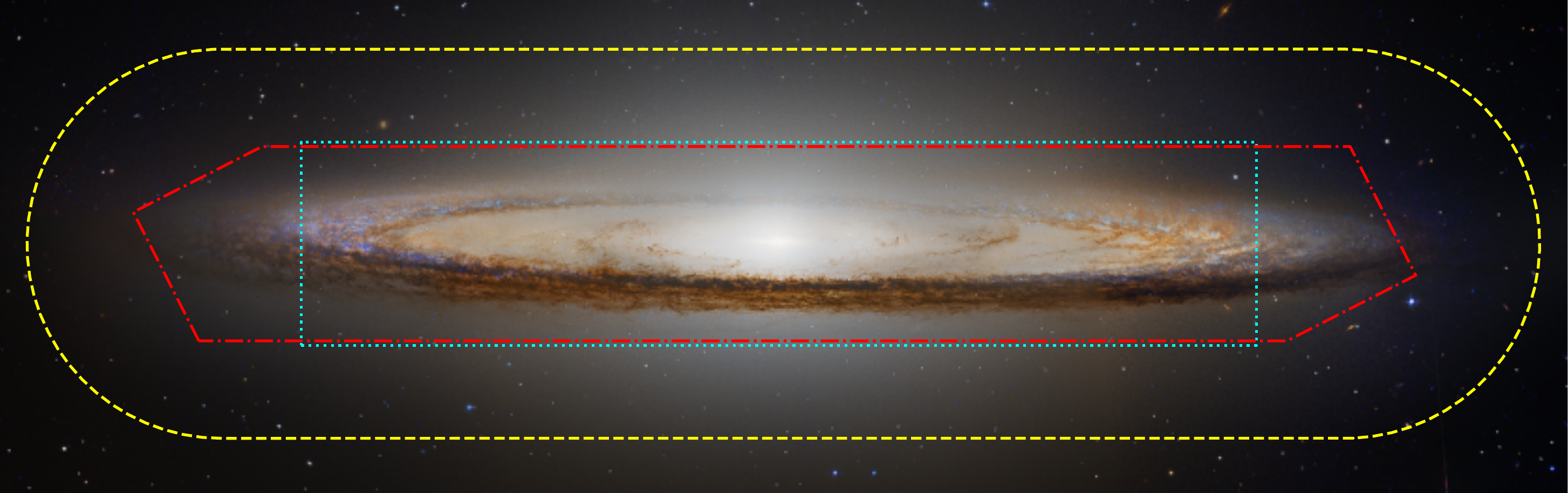}
\end{center}
\caption{Coverage of ALMA, Herschel/PACS, and MUSE data marked with yellow dashed, red dash-dotted, and cyan dotted lines, respectively, over a three color HST image of M~104 appeared as Astronomy Picture of the Day on March 29, 2019. The HST data were obtained by the Hubble Heritage Team in the filters F435W (B), F555W (V), and F625W (r).\textsuperscript{\dag}}
\footnotesize\textsuperscript{\dag}{Credits: ESA/Hubble \& NASA/R. B. Andreo,  \url{https://apod.nasa.gov/apod/ap190329.html}} 
\label{fig:coverage}
\end{figure*}

One particularly interesting galaxy for studying how unique histories and observing perspectives can effect certain diagnostics is the nearby galaxy Messier~104, commonly referred to as the `Sombrero Galaxy'. The general properties of this galaxy are summarized in Table~\ref{tab:properties}. M~104 has been classified as a SAa spiral, although there has been some debate about the exact nature of this galaxy \citep[see e.g.][and references therein]{Krause2006}.  While it appears as an edge--on spiral galaxy with a thick ring of dust, visible as a dark ring in the HST image shown in Figure~\ref{fig:coverage}, the bulge has properties similar to those of an elliptical galaxy \citep{Krause2006} and measurements of star--formation rate (SFR) and dust temperatures match more closely with early--type galaxies than similar nearby spirals.  In addition, decomposition of the structures observed at 3.6~\micron\ show that M~104 does not follow classical bulge/disk relations, and instead is better represented as an elliptical galaxy with a large halo \citep{Gadotti2012}.  

M~104 is of further interest due to the multiple forms of observational evidence indicating it hosts an active galactic nucleus (AGN).  A $10^9$M$_{\odot}$ black hole in the center of M~104 was confirmed by \citet{Kormendy1996}, a mass value further refined to $6.6\times10^8$M$_{\odot}$ by \citet{Jardel2011}.  X-ray and radio observations confirm the presence of a low--luminosity AGN with powerful jets \citep{Fabbiano1997, Pellegrini2002, Pellegrini2003, Nicholson1998, Hada2013, Vlahakis2008, Krause2006}.  Further analysis of the radio data by \citet{Bendo2006} indicate that the AGN is the dominant energy source in the central nucleus, while the outer disk is characterized by an infrared bright ring.  This complex structure and interplay between the active nucleus, unique bulge, and dusty ring makes M~104 an exciting source for studying how an edge--on early type galaxy might appear different than typical star--forming disk galaxies, providing key indicators that could identify other similar galaxies.

\begin{deluxetable}{lcc}
\label{tab:properties}
\tablecaption{General properties}
\tabcolsep=0.1cm
\tablehead{\colhead{Quantity} &\colhead{Value} &\colhead{Reference}}
\startdata
R.A.(J2000) & 12:39:59.4 & \\
Dec (J2000) &  -11:37:23 & \\
Distance & 31.1 Mpc & \citet{McQuinn2016}\\
%Redshift & (3416$\pm$17)10$^{-6}$ & \citet{Smith2000}\\
Scale  & 70 pc/arcsec & \\
Inclination & 84.3$^o\pm0.2^o$ & This work \\
v$_{sys}$ & 1089.2$\pm$1.1 km/s& [LSRK] This work\\
Type & SA(s)a & \citet{RC3}\\
\enddata
\end{deluxetable}

In order to further delve into the mysteries still surrounding M~104, we compare different gas and dust tracers from a wide--array of archival datasets. As individual emission lines and features trace different component of the interstellar medium (ISM), by compiling multi--wavelength observations of M~104 we can build a more complete picture of what is happening inside this galaxy and determine how these ISM tracers could identify similar, unresolved sources in the more distant universe.  Figure~\ref{fig:coverage} shows the footprint of the ALMA (Atacama Large Millimetre/submillimetre Array), VLT/MUSE (Very Large Telescope/Multi Unit Spectroscopic Explorer), and \textit{Herschel}/PACS (Photodetector Array Camera and Spectrometer) observations of M~104. 
One tracer of particular interest is the 157.7~\micron\ line of singly--ionized carbon (the \CII\ line).  The \CII\ line has risen in prominence as a tracer of SFR not affected by dust attenuation in distant galaxies thanks to recent surveys like ALPINE~\citep[ALMA Large Program to Investigate C+ at Early Times,][]{LeFevre2020} and REBELS~\citep[Reionization-Era Bright Emission Line Survey,][]{Fudamoto2022}.  As the wavelength of this line is in the far--infrared, it can pass through dust and gas with minimal attenuation while its role as a key cooling channel for photodissociation regions \citep[PDRs,][]{Wolfire2003} provides a physical cause for the correlation between \CII\ luminosity and SFR \citep{DeLooze2014, HerreraCamus2015}.  Unfortunately, this relationship seems to break down in the most actively star--forming galaxies \citep{DiazSantos2017}.  This is caused by an effect commonly referred to as the ``\CII\ deficit,'' which describes the decreasing trend in \CII/FIR in galaxies with increasing FIR surface brightness \citep{Sutter2022}, star--formation surface density \citep{Smith2017}, or dust temperature \citep{Croxall2012}. 

\begin{deluxetable*}{lcccclccccc}[!t]
\label{tab:ETGs}
\tabcolsep=0.1cm
\tablewidth{0pt}
\tablecaption{Early Type Galaxy Sample}
\tablehead{\colhead{Galaxy} &\colhead{RA} &\colhead{Dec} &\colhead{D} &\colhead{$z$} &\colhead{[CII] } &\colhead{[CII] Luminosity} & \colhead{$\Sigma_{FIR}$} & \colhead{UV$_{\rm{Atten}}$} &\colhead{i} & \colhead{Ref$^b$} \\ 
\colhead{} & \colhead{ J2000 } & \colhead{J2000} & \colhead{Mpc} & \colhead{} & \colhead{ObsID$^a$} & \colhead{$10^6 L_{\odot}$} & \colhead{$10^8 L_{\odot}$ kpc$^{-2}$} & \colhead{$10^6 L_{\odot}$ kpc$^{-2}$} & \colhead{deg} &\colhead{}}
\startdata
NGC~1266 & 03:16:04 & $-$02:25:40.0 & 29.9 & 0.00724 & {\it P} 1342214219 & $21.84 \pm 0.41$ & $1.93\pm 0.19$ & 82.05 & 34.8 & 1\\
NGC~3032 & 09:52:08 & +29:14:10.84 & 22.0 & 0.00517 & {\it P} 1342231734 & $8.65 \pm 0.03$ & $1.40\pm0.14$ & 23.52 & 26.4 & 2\\ 
NGC~3607 & 11:16:55 & +18:03:06.34 & 18.2 & 0.00310 & {\it P} 1342232317 & $1.48\pm0.06$ & $2.32\pm0.23$ & 12.73 & 42.4 & 3 \\
NGC~3665 & 11:24:43 & +38:45:45.82 & 22.9 & 0.00690 & {\it P} 1342234055 & $6.30\pm0.12$ & $2.13\pm0.21$ & 2.80 & 39.2 & 4\\
NGC~4459 & 12:29:00 & +13:58:41.48 & 16.5 & 0.00398 & {\it P} 1342234949 & $1.13\pm0.10$ & $2.12\pm0.21$ & 2.38 & 30.9 & 5 \\
NGC~4526 & 12:34:03 & +07:41:58.35 & 14.9 & 0.00206 & {\it P} 1342233262 & $5.13\pm0.08$ & $1.66\pm0.17$ & 0.98 & 73.8 & 5 \\
NGC~4636 & 12:42:50 & +00:10:45.07 & 15.1 & 0.00313 & {\it P} 1342236884 & $0.58\pm0.07$ & $0.56\pm0.06$ & 9.51 & 64.5 & 2 \\
NGC~4710 & 12:49:39 & +15:09:55.61 & 16.8 & 0.00368 & {\it P} 1342236273 & $10.9 \pm 0.05$ & $65.3\pm6.52$ & 2.21 & 90.0 & 6 \\
NGC~5322 & 13:49:15 & +60:11:25.92 & 21.2 & 0.00594 & {\it F} 07\_0105\_2 &$0.85\pm0.05$ & $0.07\pm 0.01$ & 29.61 & 34.5 & 7 \\
NGC~5576 & 14:21:04 & +03:16:15.6 & 25.0 & 0.00502 & {\it F} 07\_0105\_3 & $1.80\pm0.21$ & $1.06\pm0.11$ & 134.8 & 25.5 & 2 \\
NGC~5866 & 15:06:30 & +55:45:47.57 & 14.7 & 0.00252 & {\it P} 1342223744 & $4.10\pm0.28$ & $59.6\pm5.96$ & 4.91 & 90.0 & 2\\
\enddata
%\caption{Sample of Early Type Galaxies (ETGs) used a comparison sample for M104.}
\tablecomments{$a$: {\it P} marks \textit{Herschel}/PACS observations, {\it F} indicates SOFIA/FIFI-LS observations.
$b$: References for Distances: (1) \citet{Cappellari2011}, (2) \citet{Tully2013}, (3) \citet{Humphrey2009}, (4) \citet{Theureau2007}, (5) \citet{Villegas2010}, (6) \citet{Tully1988}, (7) \citet{Willick1997} }
\end{deluxetable*}

Understanding the source of this \CII\ deficit is further complicated by the ubiquity of singly--ionized carbon.  As neutral carbon has an ionization potential of 11.3~eV, well below one Rydberg, C$^+$ can exist in a variety of ISM environments spanning HII regions to giant molecular clouds (GMCs).  By investigating the \CII\ emission from nearby galaxies like M~104, we can better isolate where within the galaxy the \CII\ emission originates, allowing for further refinement of \CII--based SFR indicators.  Viewing M~104 nearly edge--on also gives us an unique perspective to study the distribution of different ISM phases, especially as we can observe the inner and outer edges of the dusty ring.  By comparing the observed \CII\ emission to indicators of unattenuated star formation like H$\alpha$ and FUV, indicators of dust heating like polycyclic aromatic hydrocarbon (PAH) emission, and indicators of molecular gas like CO, we can better determine what physical processes lead to an increase or decrease in the amount of \CII\ emission we observe from a galaxy.  This will place important constraints on how \CII\ is used to study the evolving properties of the ISM over cosmic time.  

This paper is organized as follows.  In Section~\ref{sec:data} we describe the archival datasets used to complete our analysis along with the necessary reductions used for each.  In Section~\ref{sec:results} we describe the different methods used to assess the properties of individual regions across M~104, as well as global measurements including the inclination angle and the mass of molecular gas in the ring.  Finally, in Section~\ref{sec:conclusions} we list the main findings from our analysis.

\section{Data and Observations} \label{sec:data}
\label{sec:observations}

\subsection{Herschel/PACS spectra}
\label{sec:PACS}
%Velocities are in the LSR frame.
Herschel observations of M~104 were executed as part of the ``Key Insights in Nearby Galaxies: a Far Infrared Survey with Herschel'' program \citep[KINGFISH,][]{Kennicutt2011}. Several hours were needed to scan the entire ring of dust with PACS (see the red dash-dotted line in Figure~\ref{fig:coverage}). In particular, an observation of 19,060~s (approx 5.3~hours) covered the \CII{} and [OIII]~88$\mu$m lines (obsID 1342212603) and an even longer observation of 23,988s (6.7~hours) was performed to observe the [OI]~61$\mu$m line (obsID 1342213142).
In this paper we focus our attention on the \CII{} observation which detected faint emission across the entire observed region. The oxygen lines, which are much fainter, are mainly detected in the nuclear region.

The standard mode of observation with PACS used chopping between target and sky to get rid of the varying background radiation during the observation. However, because the maximum chopper throw was 6', this method did not work for extended sources~\citep{Poglitsch2010}. In this case, the alternative ``unchopped mode" which does not involve chopping, was used~\citep{Fadda2016}.
Data reduction of this mode is known to be problematic because it is impossible to completely remove transient behaviours of the response of the detector without using off-target data taken when chopping.
Instead, transient behaviours of the response have to be modeled and removed from the data to obtain cleaner spectral cubes. As explained in more detail in the appendix \ref{sec:transients}, the case of M~104 is particularly challenging  because of the presence of a strong transient in the data which lasts for almost the entirety of the observation. In particular, at the beginning of the observation, the response is 20\% to 25\% lower than in at the end. 
Since the scan is repeated twice, data from one side of the galaxy are taken at the beginning and at the end of the observation and are combined by the standard pipeline without correcting the discrepancy in response. This generates a large gradient in the final data which is clearly visible in the spectral cubes stored in the Herschel archive. This does not only affect the level of the continuum but also the shape of the \CII{} line. For this reason, previous studies did not consider these observations, citing the poor quality of the data \citep[see, e.g., ][]{HerreraCamus2015}.

Another important point of concern is the flux calibration of the PACS data. All the data obtained in the standard spectroscopic mode of PACS (chop-nod) have been normalized to the telescope background which is extremely stable and well known during the entire mission. The unchopped data conserved in the Herschel archive instead uses the calibration blocks at the beginning of the observation for the absolute calibration. These are very sensitive to transients in the data occurring while the detector points away from sky regions with fluxes very different from the flux of the actual target. In our analysis, we use the telescope background to calibrate the data. As shown in the appendix, this introduces a correction of 50\% to 60\% with respect to the calibration obtained using the calibration blocks.

\begin{figure*}[t!]
\begin{center}
\includegraphics[width=0.8\textwidth]{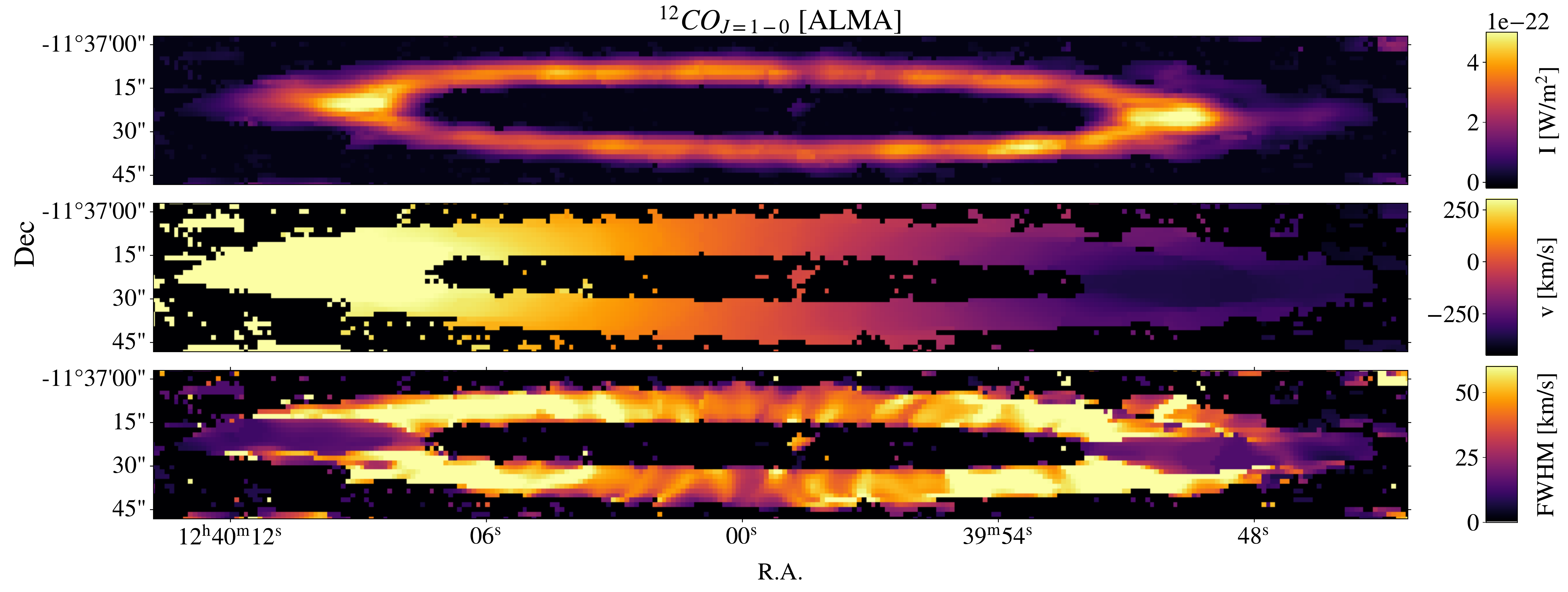}
\includegraphics[width=0.8\textwidth]{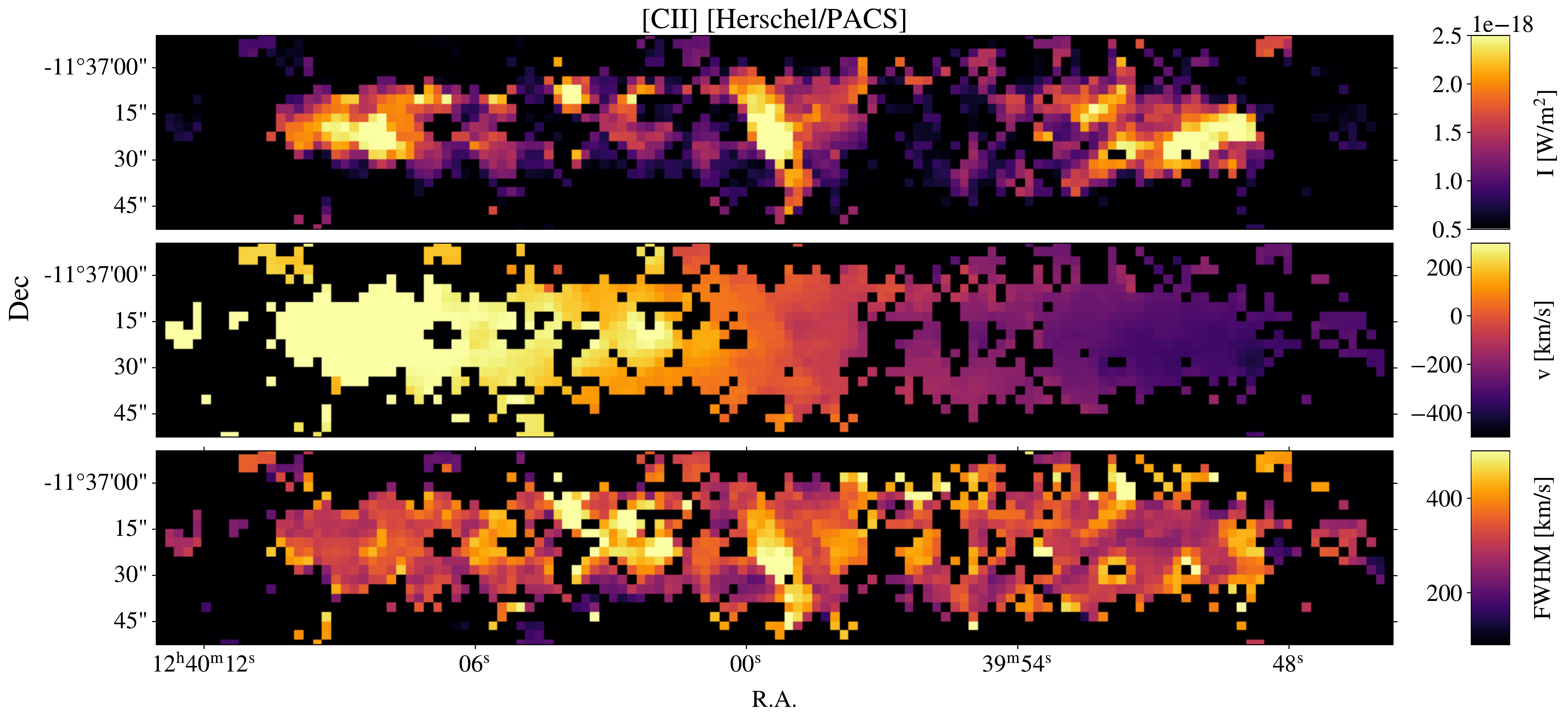}
\includegraphics[width=0.8\textwidth]{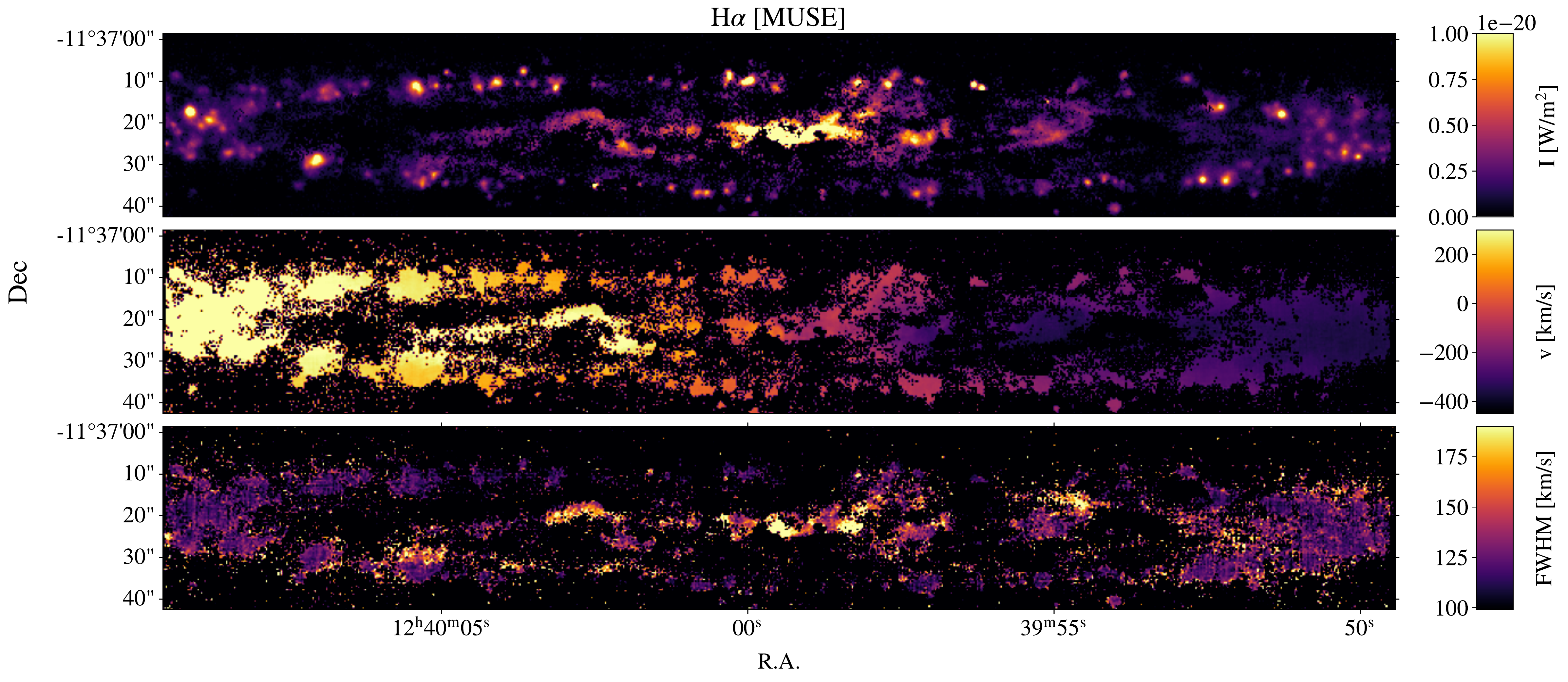}
\end{center}
\caption{Integrated intensity, velocity, and FWHM maps of the CO, \CII, and H$\alpha$ emission lines. Sky positions where lines have signal-to-noise ratios less than 5 are left black.
}
\label{fig:maps}
\end{figure*}

The data reduction has been done using the {\it unchopped pipeline with transient corrections} available in HIPE~15 \citep[see][]{Fadda2016} with some important modifications which are described in detail in the appendix~\ref{sec:transients}.

\subsection{ALMA CO data}
\label{sec:COdata}

We used unpublished archival observations of the $^{12}CO_{J=1\rightarrow0}$ line in the 2.61--2.63~mm wavelength range taken with ALMA in band~3 (program ID: 2018.1.00034.S, P.I. Catherine Vlahakis). The data were taken on 2019 March 29 and cover the whole region observed with \CII{} (see Fig.~\ref{fig:coverage}). M~104 was observed for a total of 5241~s with the 7~m compact array for an expected line sensitivity of 16.2~mJy/beam. The spatial beam is an ellipse with axes of 14\farcs3 and 7\farcs7 and major axis oriented along the major axis of the galaxy disk. The spectral resolution is 2.57~km/s. Velocities are expressed in the LSRK reference frame.
No reduction was necessary since the archival data products already had a quality sufficient for our study.

\subsection{MUSE spectroscopy}
M~104 was observed with MUSE during the science verification phase (program 60.A-9303(A)) for a total of 5.4~hours. The data quality is excellent: the median seeing during the acquisition of the data was 1.3 (with a dispersion of 0.3). The reduced data are available directly from the ESO archive as spectral cubes with 0.2~arcsec spatial pixels. However, some of the observations were coadded with incorrect offsets producing fake double sources. We decided to reprocess the 10 observations covering the region observed in [CII] and CO, shown by the blue dotted line in Figure~\ref{fig:coverage}. We used the MUSE pipeline version 2.8.5 and the pre--processed calibration data from the archive \citep{Weilbacher2020}. Data were processed through the modules {\it object} and {\it scipost}, and then combined into a single spectral cube with the module {\it exp\_combine}. 
The final cube was reprojected to a 0.4~arcsec pixel grid which roughly corresponds to a third of the spatial resolution and therefore allows a good PSF reconstruction, since the Nyquist-Shannon theorem requires a minimum of 2 pixels per resolution element. 
The wavelength calibration is done using arc lamps in standard air. The zero point is refined by using the sky lines observed in the spectrum. Small shifts are expected since the temperature of the telescope varies between the arcs and the observations by several degrees. The spectra of the different observations are shifted by less than 0.2~\AA. 

To check the quality of the wavelength calibration, we also reduced a few off-target observations obtained during the same nights in a position close to M~104 by obtaining sky spectra with the module {\it create\_sky}. We verified that the corrections were of the same order of those of the M~104 observations (around 0.2~\AA) and that the main lines close to the H$\beta$ and H$\alpha$ lines had a shift of less than 0.01~\AA.
Finally, a correction of approximately $-28.5$~km/s is applied to put the spectra into the barycentric reference frame.
ALMA and PACS spectral cubes are set to the kinematic local standard of rest (LSRK). To directly compare the cubes we changed the spectral reference system to LSRK by adding 0.9~km/s to the MUSE data (accounting for the angle between the LSRK vector and the observation direction toward M~104).

\begin{deluxetable}{lcc}
\label{tab:inclination}
\tablecaption{Inclination angle}
\renewcommand{\arraystretch}{0.87}
\tablehead{\colhead{Filter} &\colhead{Wavelength} &\colhead{Inclination} \\ 
\colhead{} & \colhead{ $\mu$m } & \colhead{Degs} } 
\startdata
IRAC Ch4 & 7.870 & 84.6$\pm$0.1 \\
MIPS 24 & 23.70  & 84.6$\pm$0.3\\
PACS 70 & 71.11 &  84.8$\pm$0.5\\
PACS 100 & 101.20 & 83.9$\pm$0.4 \\
PACS 160 & 162.70 & 84.2$\pm$0.5 \\
$^{12}$CO$_{J=1\rightarrow0}$ &2600.75 & 84.3$\pm$0.2 \\
\enddata
\end{deluxetable}

\subsection{Photometric Data}
\label{sec:phot}
Near and far-UV fluxes were obtained from archival {\it GALEX} images retrieved using the MAST archive from the Nearby Galaxy Atlas \citep[obsid 2490422561716830208,][]{https://doi.org/10.26093/cds/vizier.21730185}. For the visible bands we used SDSS data from the SDSS DR12 Science Archive Server \citep{SDSS_DR12}.

The near-IR photometry in J, H, and Ks bands was obtained from 2MASS images retrieved from the 2MASS Large Galaxy Atlas through the IRSA archive \citep{https://doi.org/10.26131/irsa122}.

The photometry between 3$\mu$m and 24$\mu$m was obtained using WISE, Spitzer/IRAC and Spitzer/MIPS (24$\mu$m) images from the IRSA archive. The WISE band 3  image at 11.3\micron\ was obtained from the all--sky survey \citep{https://doi.org/10.26131/irsa153}, while the Spitzer data were obtained from the Spitzer archive \citep{https://doi.org/10.26131/irsa413} as part of the Spitzer Infrared Nearby Galaxy Survey \citep[SINGS,][]{Kennicutt2003}.

Far-IR photometric observations were performed by the PACS instrument onboard \textit{Herschel} as part of the KINGFISH program \citep{Kennicutt2011}.  The 70~\micron, 100~\micron, and 160~\micron\ level 2 images acquired in this survey were obtained from the ESA \textit{Herschel} archive.  No further reprocessing was required for these images as the photometric measurements obtained were sufficiently accurate for our analysis.

%Far-IR photometry was based on Herschel PACS and SPIRE images were retrieved from the ESA Herschel archive \citep{Kennicutt2011}. All of these images did not need any reprocessing since the photometric measurements obtained were sufficiently accurate for our analysis.

\subsection{ETG comparison sample}
\label{sec:etgs}
Although M~104 is classified as an SAa galaxy by \citet{RC3}, the classification is uncertain given the high inclination which does not allow us to discern the features of its disk. A study of the kinematical properties of its globular clusters \citep{Dowell2014} found that M~104 is has similar properties to early-type galaxies. In this paper we consider a set of early-type galaxies (ETGs, in the following), which have been observed either with PACS on Herschel or with FIFI-LS on SOFIA as a comparison set.  Specifically, the data from these galaxies are included in Sections~\ref{sec:CIIUV} and \ref{sec:CIIFIR}, where the \CII\ fluxes are compared to the amount of attenuated UV light and the FIR luminosities.

\begin{figure*}[!t]
\begin{center}
\includegraphics[width=0.75\textwidth]{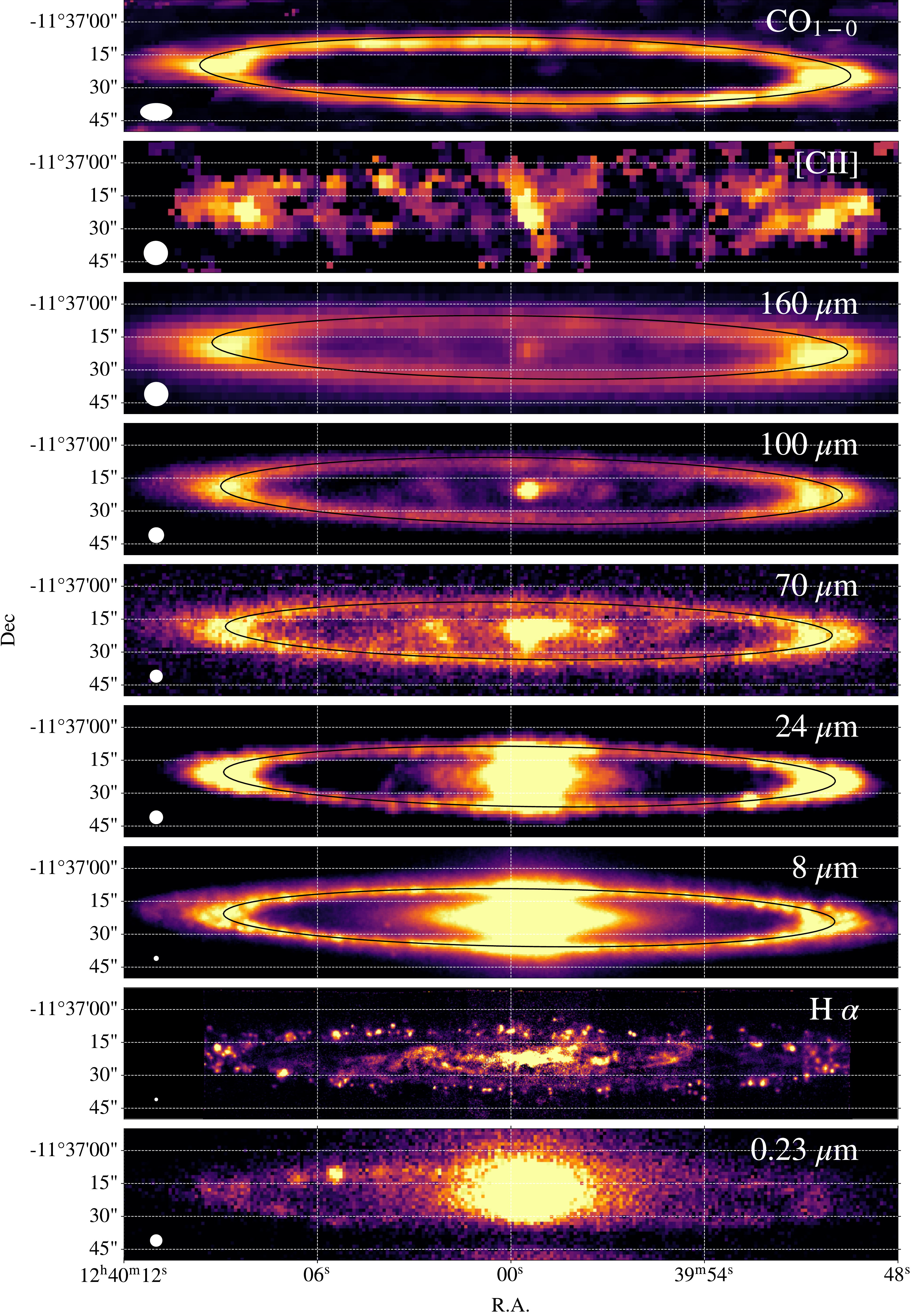}
\end{center}
\caption{The ring of the Sombrero galaxy as seen from UV to sub-millimeter wavelengths. The PSF of each image is shown in the left bottom corner of each panel. The ellipse fitted to the ring is marked in black. NUV, H$\alpha$, and [CII] intensity images only have bright spots along the ring and weak continuous component.
}
\label{fig:inclination}
\end{figure*}

The selected galaxies have a complete set of optical and near- to far-IR observations to perform the same analysis used with M~104 when comparing their total infrared and [CII] emission.
The list of galaxies is given in Table~\ref{tab:ETGs}.  All values listed are global measurements, as individual regions in these ETGs were not resolved in the available data.

\subsection{Additional Reference Samples}
In addition to the ETGs described above, we compare M~104 to nearby, star--forming galaxies which have similar data available in the literature.  As there are limited local galaxies with measurements of \CII, CO, FIR luminosity, and PAH emission of similar quality and resolution as the M~104 data described previously, this sample is limited and not all the galaxies described here can be included in the full analysis.  Primarily, we include data from local star--forming galaxies NGC~7331 \citep{Sutter2022}, NGC~1097 and NGC~4559 \citep{Croxall2012}, M~31 \citep{Kapala2017}, as well as the KINGFISH survey \citep{Kennicutt2011}.    In our comparison between \CII\ and FIR we also include data from the Great Observatories All sky LIRG Survey \citep[GOALS][]{DiazSantos2017} and a survey of $z \sim 0.02 - 2$ galaxies from \citet{Ibar2015} to provide further context for our data.

\section{Results and Discussion}
\label{sec:results}

\subsection{CO, \CII, and \texorpdfstring{H$\alpha$}{} maps}
\label{sec:maps}

We fit the CO, \CII, and H$\alpha$ lines in the ALMA, PACS, and MUSE spectral cubes to produce maps of the line intensity, velocity, and velocity dispersion using the interactive software {\it sospex}~\citep{Fadda2018}\footnote{\url{https://github.com/darioflute/sospex}}. The results are displayed in Figure~\ref{fig:maps}. The pixels where the signal-to-noise ratio of the line fit was lower than 5 are left black.
In all the maps is evident that the ring of molecular gas and dust is well defined. However, while the CO map shows almost uniform emission along the entire ring, the \CII{} and H$\alpha$ emissions originate from multiple individual knots. The regions with peaks of H$\alpha$ emission are also traced in the \CII{} map, taking into account the different spatial resolution. 
The apparent higher intensity at the extremes of the ring is due to a projection effect. In fact, since the galaxy has an high inclination angle, we see a larger portion of the ring along the line of sight at the two ends of the ellipse (along the major axis) than in the other regions.
\begin{figure*}[!t]
\includegraphics[width=0.99\textwidth]{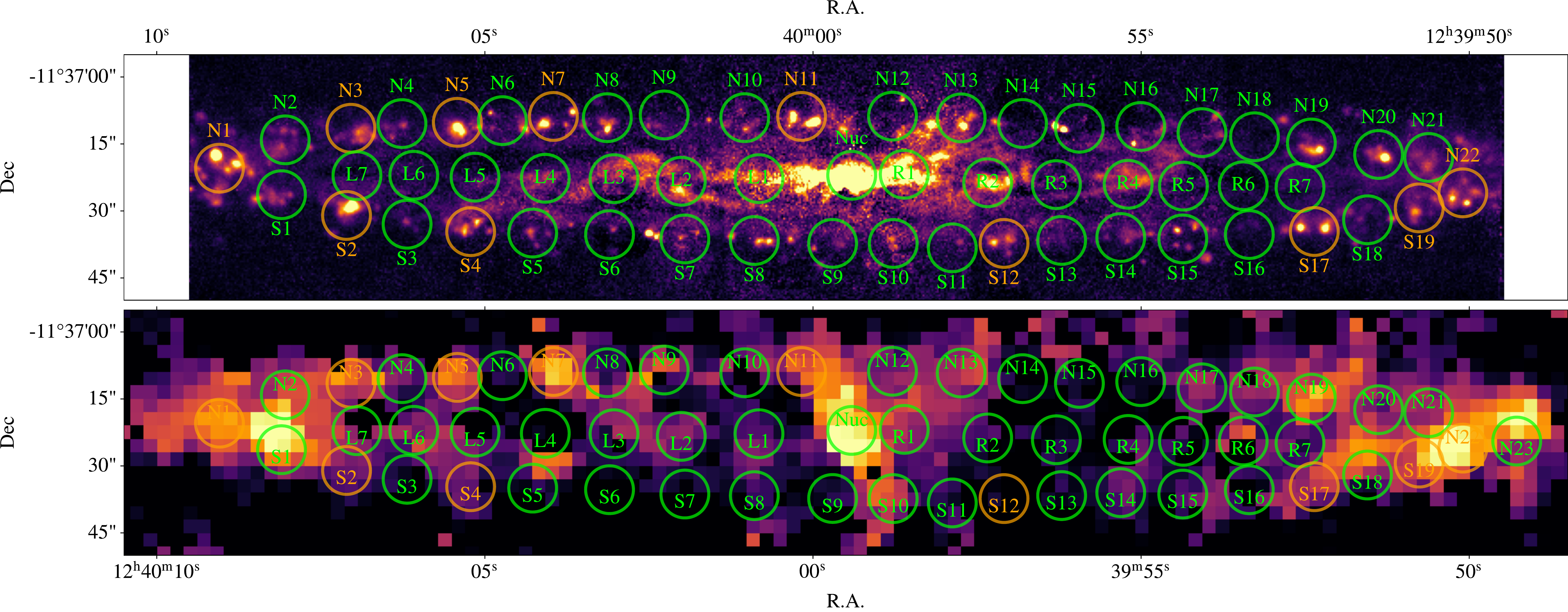}
\caption{Circular apertures selected for the detailed study of the different parts of the ring and the inner regions plotted over the H$\alpha$ and \CII{} images in the top and bottom panels, respectively. The independent apertures uniformely sample the ring and are usually centered on star forming knots. The aperture diameter corresponds to the \CII{} beam. The circles are orange or green if the region is classified as HII or LIER according to the diagnostic diagrams (see Section~\ref{sec:BPT}).
}
\label{fig:apertures}
\end{figure*}

The velocity map has a very well defined gradient across the galaxy. In particular, taking into account the projection, the angular velocity along the ring is constant (see the analysis in section~\ref{sec:kinematics}).
The map of the velocity dispersion is rather interesting. It appears that the CO ring has a very low velocity dispersion at the two extremes of the major axis of the ellipse. By analyzing the spectra (see examples in Fig.~\ref{fig:examples}), we can see that at the two ends of the major axis of the ellipse we see a single component of the molecular gas emission. Therefore, the velocity dispersion is very low. The vertical extension of the region with low velocity dispersion perfectly corresponds to the beam of the ALMA observation. As we consider parts of the ring in other angles, the beam intercepts several molecular clouds with slightly different velocity. These components are very close in velocity and difficult to resolve. For this reason, the profile of the line is approximately Gaussian and its velocity dispersion increases dramatically along the rest of the ring. The minima of the velocity dispersion are visible in the parts of the ring along the minor axis of the ellipse where, because of the projection angle, the smallest arc of the circular ring is covered by the ALMA beam.
In the case of the {\it Herschel} observations, the velocity resolution of the \CII{} line is not sufficient to distinguish different components along the ring. So, the velocity dispersion map appears rather uniform. The velocity dispersion peaks on the nucleus of the galaxy and in a few regions inside the ring.

The H$\alpha$ observations have a spectral resolution better than {\it Herschel} and enough spatial resolution to detect single regions with peaks of star formation. For this reason, MUSE  does not see the same projection effects as ALMA and the velocity dispersion is quite uniform along the ring.
In the inner region, parts of the bar have higher velocity dispersion. The velocity dispersion peaks in the nuclear and circumnuclear regions.

\subsection{Galaxy inclination angle}
\label{sec:inclination}

An important property needed to correctly compute the surface brightness of a spiral galaxy is the inclination of the disk. In literature, the inclination of M~104 has been measured at values between 84$^o$\citep{Emsellem1996, Bendo2006} and 90$^o$\citep{Jardel2011}. 
Figure~\ref{fig:inclination} shows the appearance of the ring at different wavelengths from the GALEX observations in the UV to the CO observations with ALMA. It is clear that the ring is very well defined by its dust emission, while the emission from the nucleus becomes much less dominant when moving into the far-IR. Taking advantage of the clarity with which the ring of the Sombrero galaxy is visible in infrared and millimeter images, it is possible to fit an ellipse and compute a more accurate value of the inclination, based on the hypothesis of a circular ring.
We found that the ring is perfectly fit by an ellipse once the central bulge is masked using the direct linear least squares fitting method as described by \citet{Halir1998}.

We report the results of the fit in the different images in Table~\ref{tab:inclination}. The errors reported have been computed by using different flux thresholds to select the pixels used in the fitting of the ellipses and can be considered an estimate of the stability of the fit.
The values are within a small range (0.7$^o$). Since the images which are less disturbed by the nuclear emission are the PACS 160$\mu$m and the ALMA CO intensity image, we accept the value fitted in the case of the CO: 84.3$^o\pm0.2^o$. This value is in very good agreement with the ones reported by \citet{Emsellem1996} and \citet{Bendo2006}.

\subsection{Aperture spectra}
\label{sec:apertures}

To study the different parts of the ring and its interior we defined a series of apertures. Such apertures have the diameter of the \CII{} beam (10.7~arcsec). We placed them all along the ring usually centered on star forming regions, locations of which were determined using the peaks of the \CII{} and H$\alpha$ images (see Fig.~\ref{fig:apertures}). For this reason, some of the aperture have very faint H$\alpha$ counterparts or some apertures are not centered on the point source in the H$\alpha$ image (see, e.g., N7).
In the region inside the ring, we placed an aperture on the position of the nucleus and a series of apertures on both sides of it. 

For each of these apertures we extracted spectra from the MUSE, ALMA, and Herschel spectral cubes.  The same circular apertures have been also used to define the spectral energy densities (SEDs) to derive the total infrared flux (see Section~\ref{sec:SED}). In the case of ALMA we used apertures with the same center but with the elliptical shape of the ALMA beam.

\begin{figure}[!b]
\begin{center}
\includegraphics[width=0.48\textwidth]{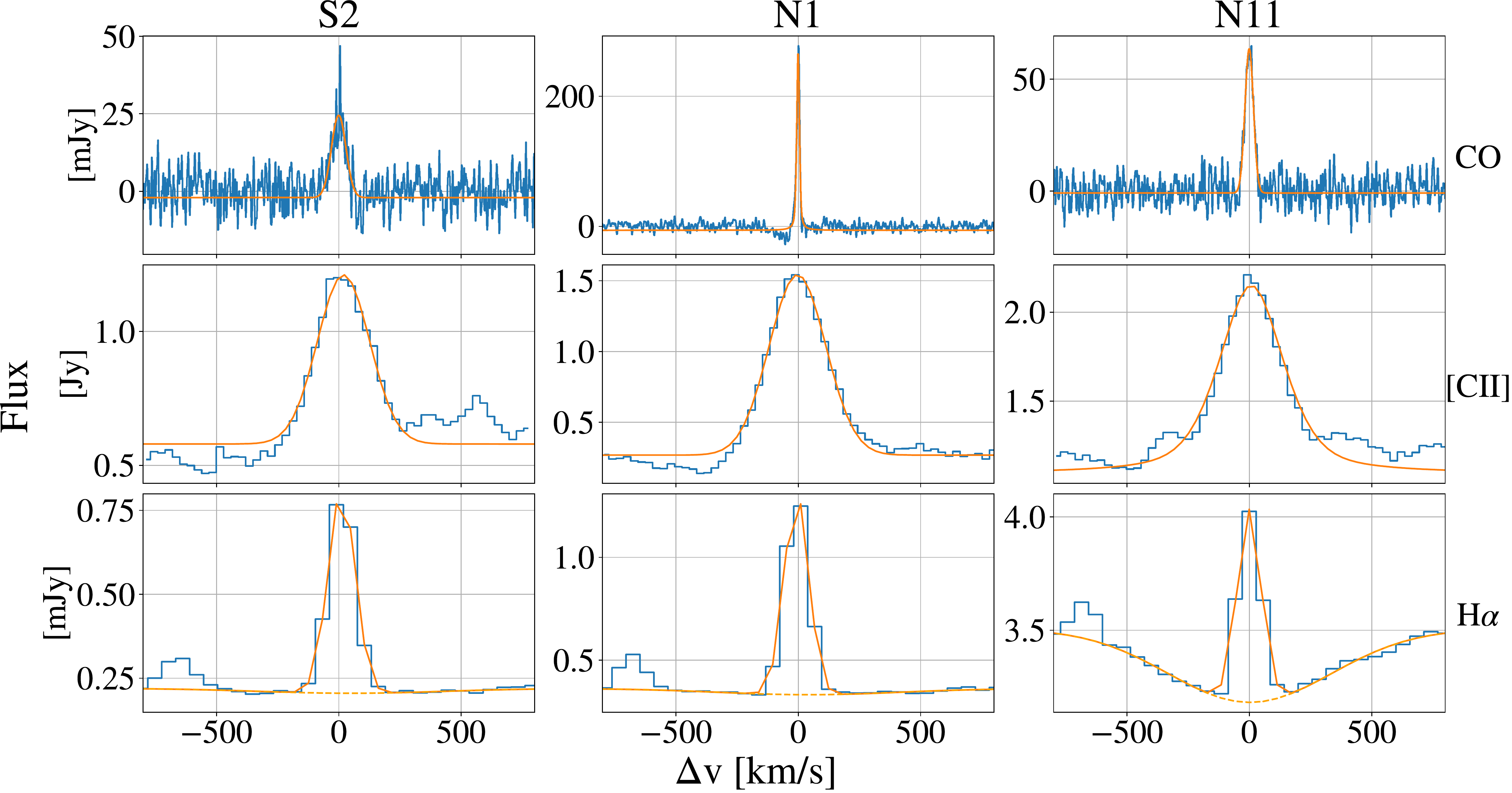}
\end{center}
\caption{
Examples of spectra in three apertures. The CO, \CII, and H$\alpha$ lines are shown in the top, middle, and lower panels, respectively. The orange line corresponds to the adopted fit.
}
\label{fig:examples}
\end{figure}

In Fig.~\ref{fig:examples} we show a few significant examples of spectra extracted from three regions in the ring. We notice in particular that some of the optical spectra show signatures of high absorption, such as the N11 aperture (spectra in the right column). In many cases, this strong absorption makes it impossible to see the H$\beta$ line in emission. The H$\alpha$ and surrounding [NII] lines are visible in all the extracted spectra. 
The example of aperture S2 shows a case where CO has probably multiple components, while the N1 aperture capture a single component along the line of sight for the CO line which appears intense and with very low velocity dispersion. Finally, N11 is situated almost in the direction of center of the ring. Since the aperture covers an arc on the ring much smaller than that of S2, it includes less components and appears to have a velocity dispersion intermediate between N1 and S2.

We fit the CO and \CII{} spectra using {\it sospex}~\citep{Fadda2018} while for the 
MUSE spectra we used the software {\it showspectra}~\footnote{\url{https://github.com/darioflute/showspectra}} which has been developed for the analysis of optical spectra.

\subsection{Kinematics of stars and gas}
\label{sec:kinematics}

\begin{deluxetable}{lccccl}
\label{tab:sedbands}
\renewcommand{\arraystretch}{1.0}
\tablecaption{Bands used for SED Fitting}
\tablehead{\colhead{Filter} &\colhead{Wavelength}&\colhead{Beam}&\colhead{Pixel}&\colhead{$\sigma_{\rm{cal}}$} & \colhead{Refs} \\ 
\colhead{} & \colhead{ $\mu$m } &\colhead{arcsec}&\colhead{arcsec}& \colhead{} & \colhead{}} 
\startdata
GALEX\_FUV & 0.152 &4.2&1.5& 0.05 mag & 1\\
GALEX\_NUV & 0.227 &5.3&1.5& 0.03 mag & 1 \\
SDSS\_$u$  & 0.354 &1.4&0.4& 2\% & 2 \\
SDSS\_$g$  & 0.477 &1.4&0.4& 2\% & 2 \\
SDSS\_$r$  & 0.632 &1.4&0.4& 2\% & 2 \\
SDSS\_$i$  & 0.762 &1.4&0.4& 2\% & 2 \\
SDSS\_$z$  & 0.913 &1.4&0.4& 2\% & 2 \\
2MASS\_$J$ & 1.235 &2.9&2.0& 0.03 mag & 3 \\
2MASS\_$H$ & 1.662 &2.8&2.0& 0.03 mag & 3 \\
2MASS\_$Ks$& 2.159 &2.9&2.0& 0.03 mag & 3 \\
IRAC\_1    & 3.550 &1.66&1.2& 1.8\% & 4 \\
IRAC\_2    & 4.490 &1.72&1.2& 1.9\% & 4 \\
IRAC\_3    & 5.730 &1.88&1.2& 2.0\% & 4 \\
IRAC\_4    & 7.870 &1.98&1.2& 2.1\% & 4 \\
WISE\_3    & 11.56 &6.5&2.75& 4.5\% & 5\\
MIPS\_24   & 23.70 &4.9&2.5& 4.0\% & 6\\
PACS\_70   & 71.11 &5.6&3.2& 5\% & 7 \\
PACS\_100  & 101.20&6.8&3.2& 5\% & 7 \\
PACS\_160  & 162.70&10.7&6.4& 5\% & 7 \\
%SPIRE\_250 & 249.40 & 4\% & 8 \\
\enddata
\tablecomments{Beam sizes for SDSS and 2MASS are median seeing values. References: (1) \citet{Morrissey2007}, (2) \citet{Padmanabhan2008}, (3) \citet{Skrutskie2006}, (4) \citet{Reach2005}, (5) \citet{Jarrett2011}, (6) \citet{Engelbracht2007}, (7) \citet{Balog2013}}
\end{deluxetable}

\begin{figure}[t!]
\begin{center}
\includegraphics[width=0.49\textwidth]{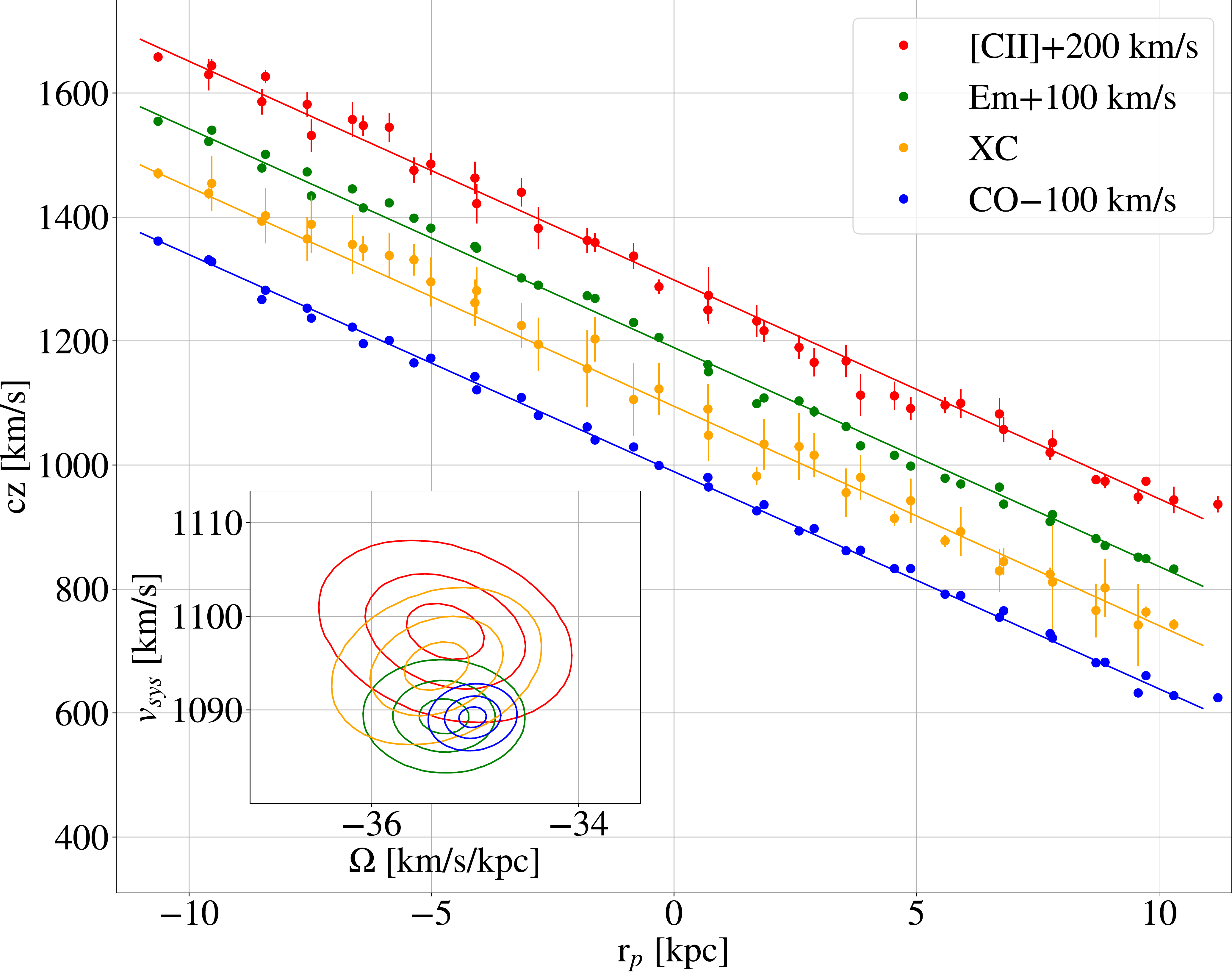}
\end{center}
\caption{ Line-of-sight velocity along the ring for the optical, \CII, and CO emission as a function of the projected distance from the galaxy center. The error bars correspond to 3-sigma errors. The different measurement sets have been spaced for clarity by 100~km/s intervals. In the legend, XC and EM stand for cross-correlation and emission lines, the two methods used to estimate redshifts from the optical spectra. The inset displays the ellipses of confidence for the 4 linear fits at the 1, 2, and 3-sigma levels \citep[39\%, 86\%, and 99\% level of confidence for two variables, see e.g.][]{Wang2015}.
}
\label{fig:veldiagram}
\end{figure}

\begin{deluxetable}{lc|c}
\tablecaption{Parameter values for CIGALE modules}
\label{tab:sedprops}
\renewcommand{\arraystretch}{1.0}
\tabcolsep=0.2cm
\tablehead{\colhead{Name} & \multicolumn{2}{c}{Values}}
\startdata
\hline
& \multicolumn{2}{c}{SFH Delayed}\\
%\hline
\texttt{tau\_main} [Tyr] & \multicolumn{2}{c}{0.5, 1.1, 1.7, 2.6, 3.9, 8.8, 13, 20}\\
\texttt{age\_main} [Tyr] & \multicolumn{2}{c}{4.5, 7, 9.5, 12} \\
\texttt{f\_burst} & \multicolumn{2}{c}{0.0}\\
\hline
& \multicolumn{2}{c}{SSP }\\
Model & \multicolumn{2}{c}{\citet{BruzualCharlot2003}}\\
%\hline
\texttt{imf} & \multicolumn{2}{c}{1 (Chabrier)}\\
\hline
& \multicolumn{2}{c}{Nebular Emission}\\
%\hline
\texttt{logU} & \multicolumn{2}{c}{-3.0} \\
\hline
&\multicolumn{2}{c}{Dust Attenuation}\\
Model &\multicolumn{2}{c}{\citet{Calzetti2000}}\\
%\hline
\texttt{E\_BV\_nebular} [mag] &\multicolumn{2}{c}{0, 0.005, 0.0075,} \\
 & \multicolumn{2}{c}{ 0.011, 0.026, 0.058,  } \\
 & \multicolumn{2}{c}{ 0.2,  0.44, 0.66, 1 } \\
\texttt{uv\_bump\_amplitude} & \multicolumn{2}{c}{0, 1.5, 3 (Milky Way)} \\
\texttt{powerlaw\_slope} & \multicolumn{2}{c}{[-0.5, 0], $\delta=0.1$}\\
\hline
\hline
\colhead{Name} & \colhead{Inner} & \colhead{Ring}\\
\hline
& \multicolumn{2}{c}{Dust Emission}\\
Model & \multicolumn{2}{c}{\citet{Draine2014}}\\
%\hline
\texttt{qpah} & 0.5, 1.1 & 0.5, 2.5, 4.6, 6.6 \\
\texttt{umin} & 0.1, 0.3,   & 0.1, 0.5,  \\
 &  0.5, 1.2, 2  & 1, 2.5, 5 \\
\texttt{gamma} & 0, 0.001,  & 0.001, 0.002,\\
               &  0.008,0.03, & 0.01,  0.016, 0.03 \\
               &  0.13, 0.25, 0.5&   0.06, 0.13\\
\hline
& \multicolumn{2}{c}{AGN Emission}\\
Model & \multicolumn{2}{c}{\citet{Fritz2006}}\\
%\hline
\texttt{f\_AGN} & 0.1, 0.25, 0.5 & 0 \\
\texttt{psy} & 0.001, 50, 90 & 50 \\
\hline
\enddata
\tablecomments{ Multiple values for a parameter define a grid of models.
Values of unlisted parameters are left at the CIGALE defaults.}
\end{deluxetable}

CO, \CII{}, and optical spectra lines give information about the kinematics of the different components of the ring. In particular, the CO emission is directly related to molecular gas, \CII{} is mainly emitted by molecular and atomic gas, the H$\alpha$ emission line is linked to ionized gas, and the absorption lines in the optical spectra are related to stars \citep[see, e.g., ][]{Sofue2001}.
The kinematics of these different components can be visualized by plotting the line-of-sight velocities as a function of the projected distance from the center of the galaxy. In fact, assuming a circular ring rotating with the angular velocity $\Omega$, the component of the velocity along the line of sight is:
\begin{equation}
    v_{los} = r_{0}  \Omega  \cos\phi + v_{sys} = r_{p}  \Omega + v_{sys},
\end{equation}
where $r_0$ is the radius of the ring in kpc, $r_{p}$ is the projected distance from the galaxy center of a point in the ring at the angle $\phi$, and $v_{sys}$ is the systemic velocity.
We therefore expect a linear relationship between the velocity along the line of sight and the projected distance from the galaxy center with slope equal to the angular velocity $\Omega$ and intercept corresponding to the systemic velocity $v_{sys}$.

For each component we used a different technique to obtain accurate measurements of velocities.

In the case of the the ALMA and {\it Herschel} data, we fit the CO and \CII{} lines using {\it sospex}. We fit the \CII{} line using a Voigt function which accurately describes the profile of the line. In the case of the ALMA data we opted for fitting Gaussian functions, since the apertures across the ring intercept several molecular clouds with slightly different velocities, but these are unresolved.  This observational effect broadens the emission line and makes the Gaussian model a better fit for the CO data.

In the case of the MUSE data, we measured the redshifts from the optical spectra of the selected apertures using two methods: cross-correlation with SDSS templates (mainly based on absorption features) and measurement of emission lines (mainly H$\alpha$, [NII], [SII], and occasionally [OIII]5007\AA\ and H$\beta$). Since the SDSS templates are expressed in vacuum wavelengths and the MUSE spectra in air wavelengths, we converted the MUSE spectra to vacuum wavelengths for the cross-correlation using the IAU standard conversion specified in \citet{Morton1991}. The redshift errors for the cross-correlation estimates were computed as in \citet{Tonry1979}. The redshifts based on the emission lines were computed through the mean weighted with the fitting errors of the lines.

Figure~\ref{fig:veldiagram} shows the line-of-sight velocities of the spectra in the different apertures along the ring as a function of their projected distance from the galaxy center. The four different colors refer to measurements of the CO line, \CII\ line, and the two ways of measuring the redshift of the optical spectra: EM for emission lines, and XC for cross-correlation. 
To better show the linear fit of the different set of points we spaced them by 100~km/s intervals. The data of the \CII{}, CO, and optical spectra give compatible values for slope and systemic velocity inside the errors. In particular, the measurements with the lowest errors (CO lines and optical emission lines) are almost identical.
The systemic velocity from the optical emission line is $(1089.1\pm1.8)$~km/s, while the one based on CO data is $(1089.2\pm1.1)$~km/s. These values agree very well with the measurement from \citet{Lavaux2011} who reported a systemic velocity of $(1090\pm5)$~km/s.
The four different methods also provide very similar values for the slope of the line, which corresponds to the angular velocity of the galaxy rotation $\Omega$. The absolute values of the angular velocity for the CO data and the optical emission lines, which have the smallest errors, are $(35.0\pm0.1)$~km~s$^{-1}$kpc$^{-1}$ and $(35.2\pm0.2)$~km~s$^{-1}$kpc$^{-1}$, respectively. Therefore, we have an almost perfect agreement between the kinematic of the stars and that of the interstellar medium. This is not always the case since in several galaxies these two components have different rotational patterns \citep[see, e.g.][]{Williams2021, Su2022}.

\begin{figure}
\begin{center}
\includegraphics[width=0.39\textwidth]{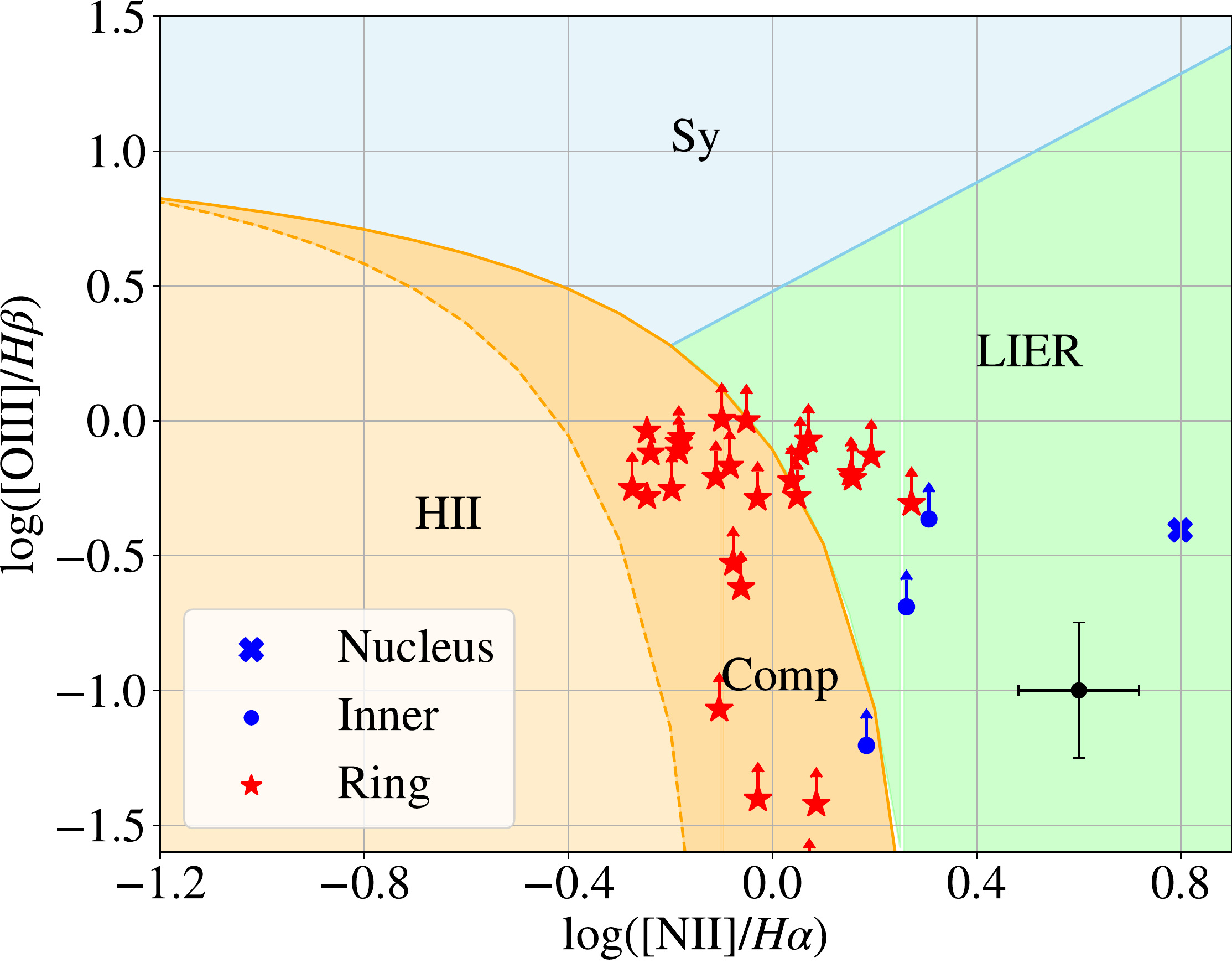}
\includegraphics[width=0.39\textwidth]{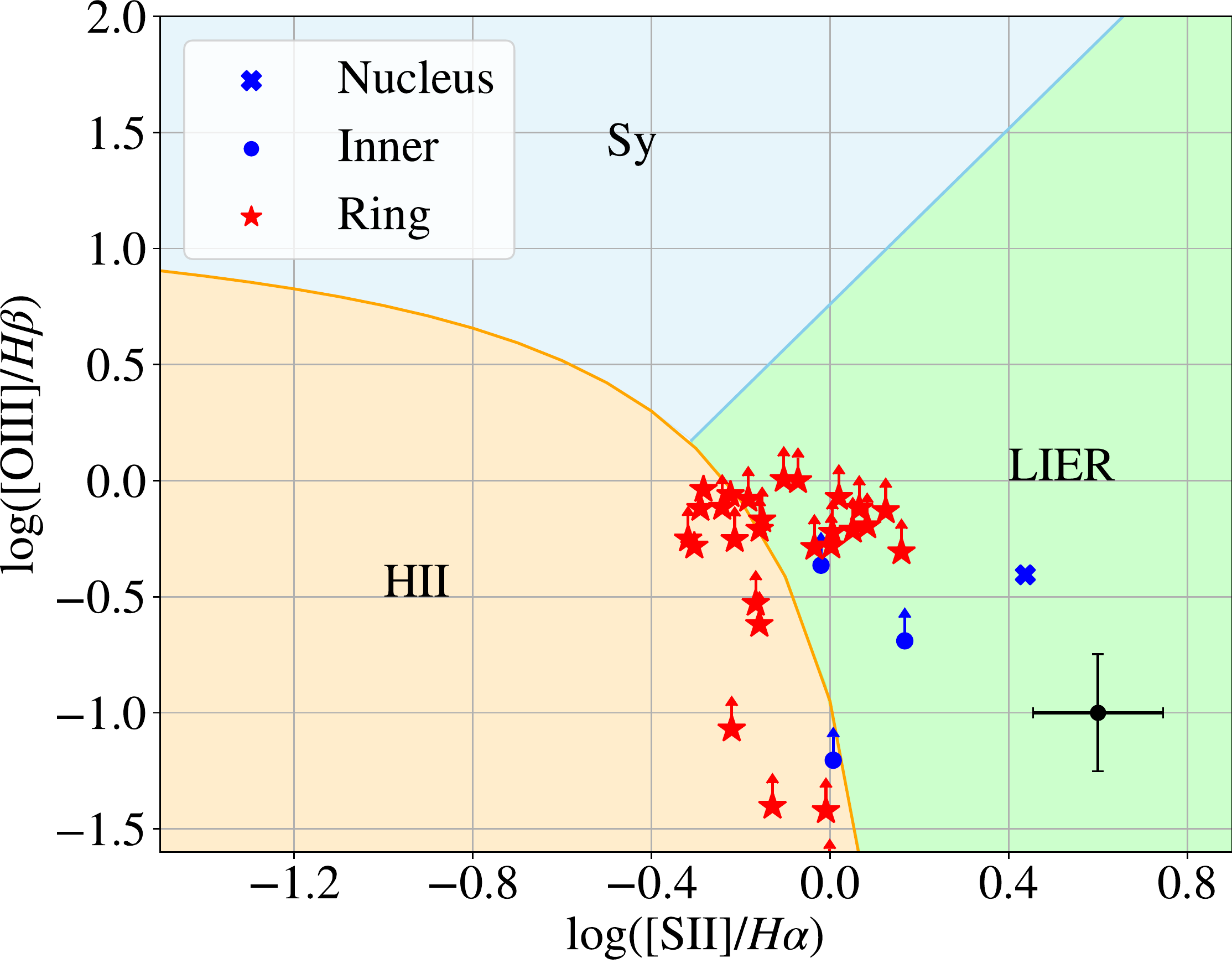}
\includegraphics[width=0.39\textwidth]{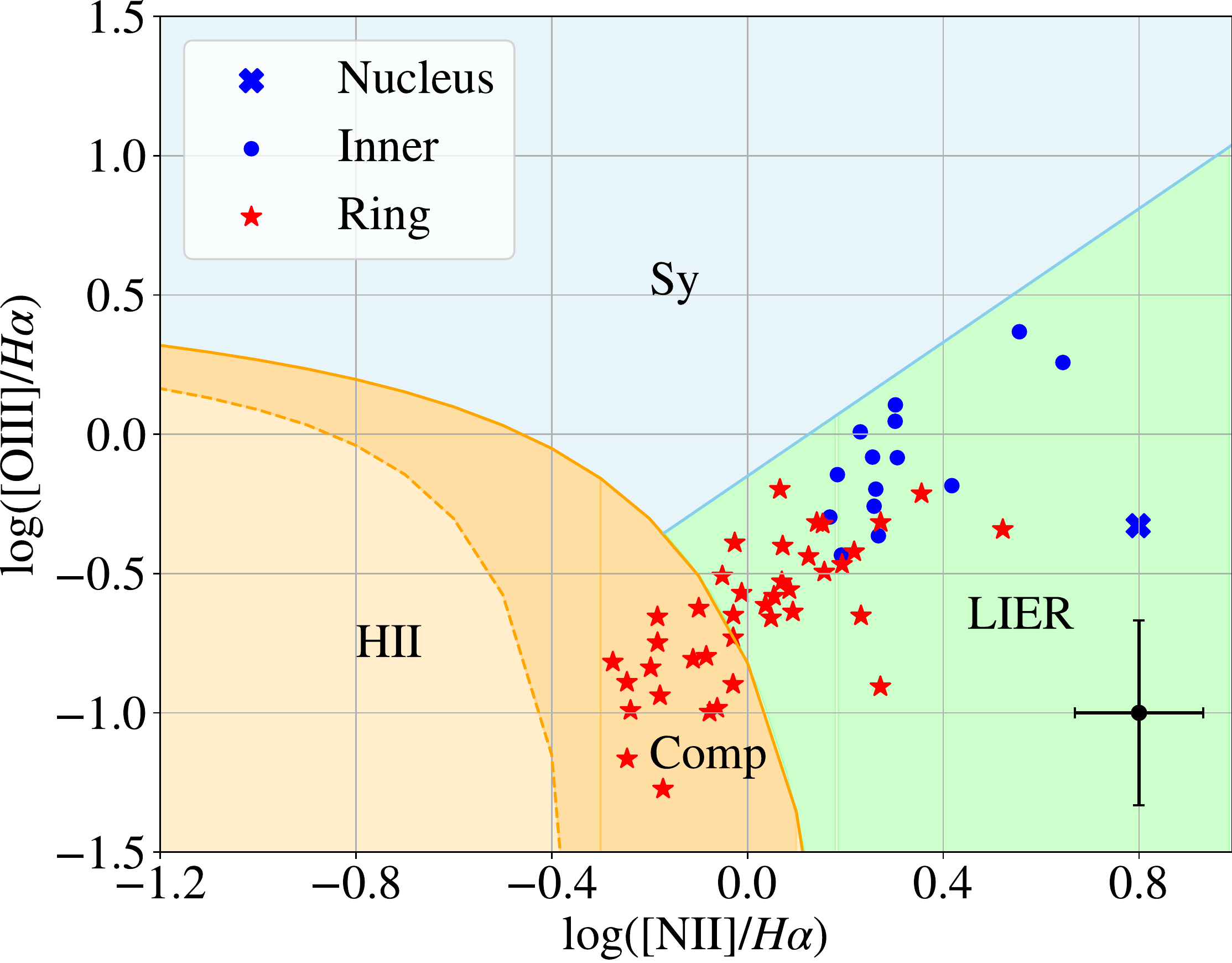}
\includegraphics[width=0.39\textwidth]{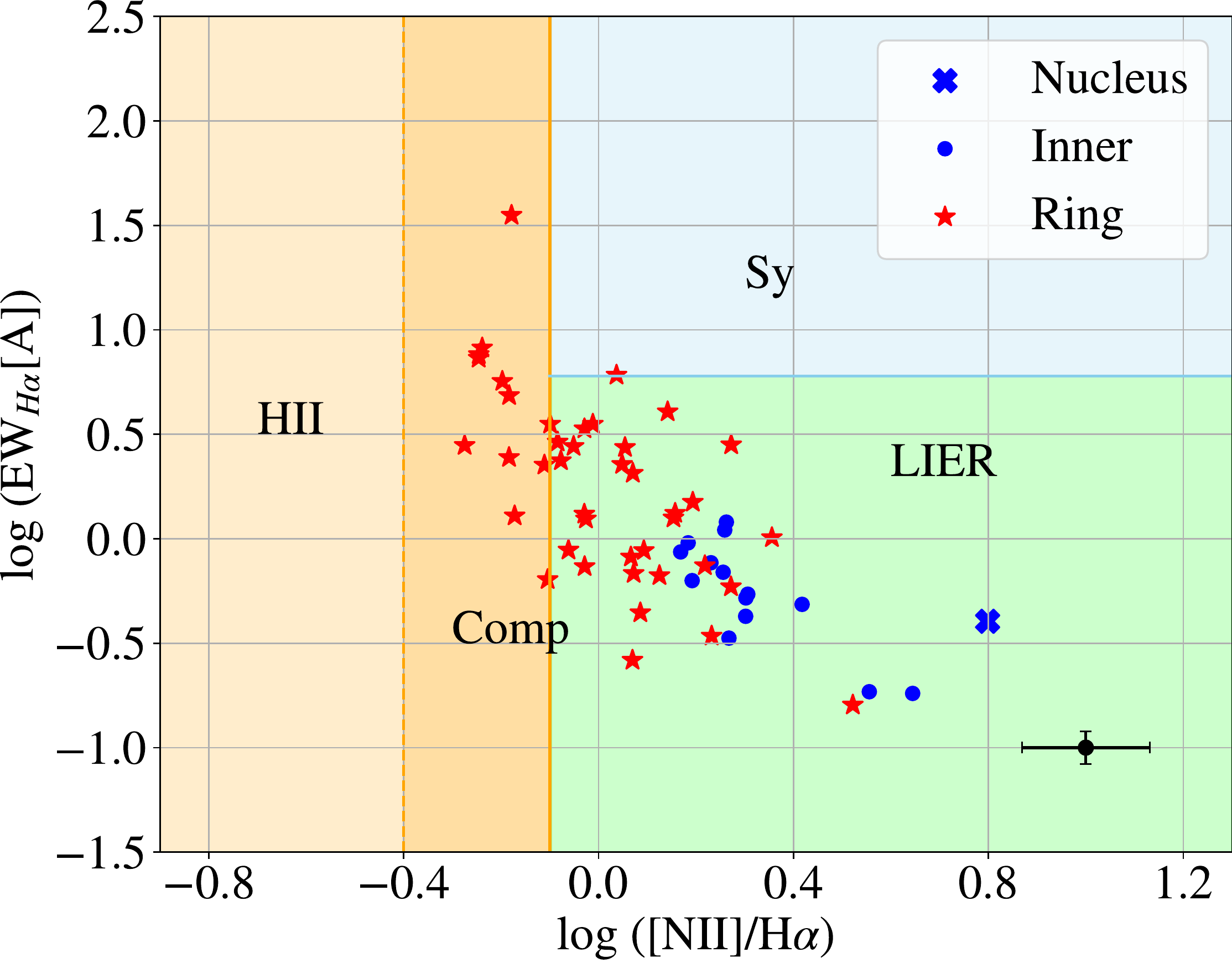}
\end{center}
\caption{Emission line diagnostic diagrams used to identify LIER and star-forming regions. The regions are labelled as Seyfert (Sy), HII, Composite (Comp), and LIER. Median errorbars are shown in the right bottom corner.}
\label{fig:BPT}
\end{figure}

\subsection{LIER and HII regions}
\label{sec:BPT}

The spectra extracted from the MUSE spectral cube for the apertures defined in Fig.~\ref{fig:apertures} have been also used to study the properties of the respective regions. Many spectra, especially those inside the ring, show heavy absorption. The Balmer lines (H$\alpha$ and H$\beta$) were always fit using a component in absorption and one in emission. H$\alpha$, [NII], and [SII] lines are visible in all the spectra. In contrast, due to attenuation, the [OIII] and H$\beta$ lines are not always measurable in emission. Figure~\ref{fig:BPT} shows several diagnostic diagrams obtained using the measurements of the emission lines. The typical diagnostic diagrams which make use of the H$\beta$ emission line are not possible for many of the spectra. In the cases where this line is barely visible, we show 3-$\sigma$ lower limits. To determine the classification for all the defined apertures, we used two alternative diagnostic diagrams developed for weak line galaxies as described in \citet{CidFernandes2010} which are shown in the two bottom panels of Fig.~\ref{fig:BPT}.
Following the example of \citet{Belfiore2016}, we named the region shaded in green as LIER (low-ionization emission-line region). The ionizing photons in such regions were originally attributed to central, low-luminosity active galactic nuclei and called LINERs (low-ionization nuclear emission-line regions). However, in the case of passively-evolving galaxies with old stellar populations, hot post-asymptotic giant branch stars are now considered to be the main source of ionizing photons, especially when outside of the nuclear region.

\begin{figure}
\begin{center}
\includegraphics[width=0.49\textwidth]{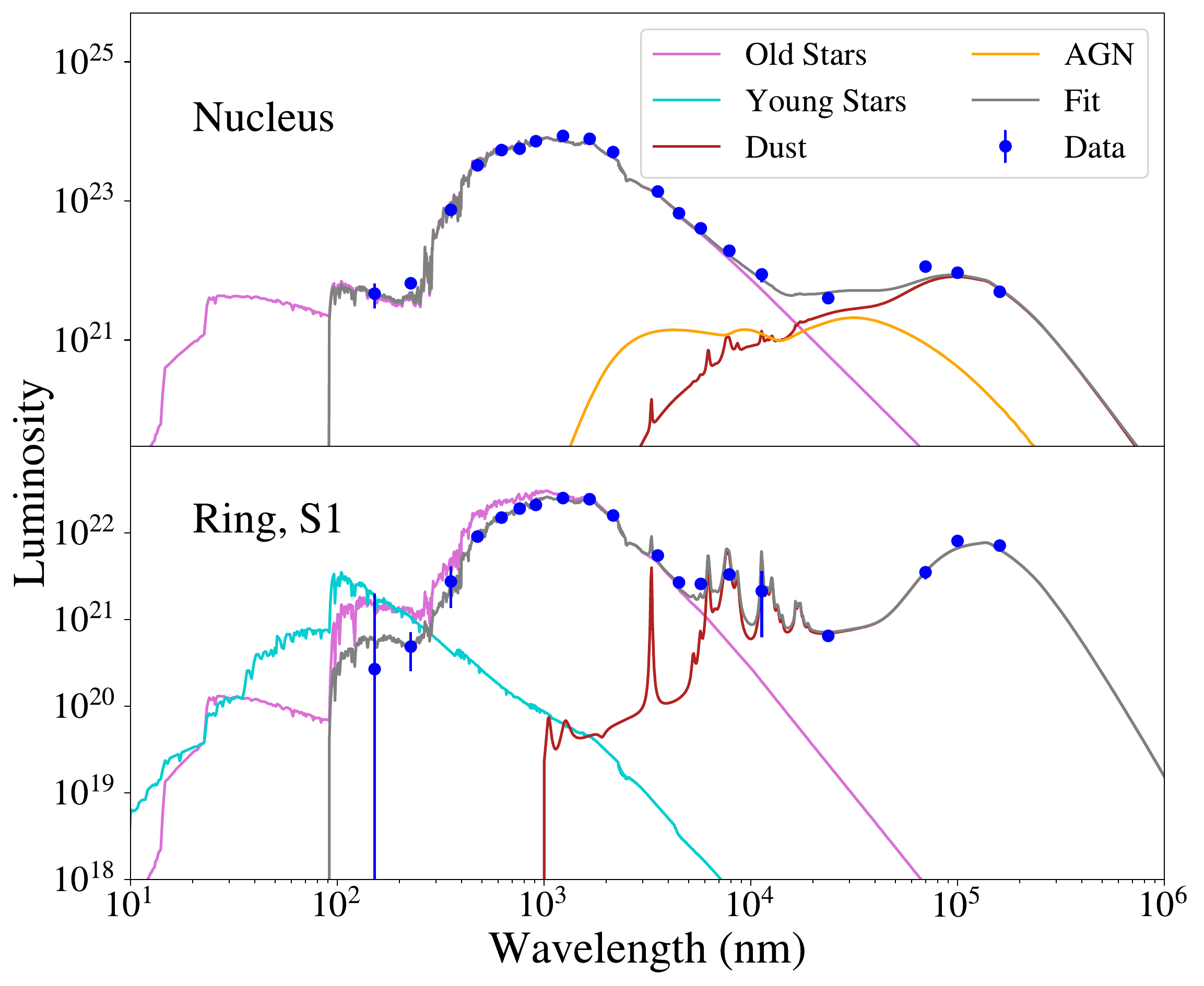}
\end{center}
\caption{
Two examples of the spectral energy distribution fits produced by CIGALE: nucleus (top) and the S1 region on the ring (bottom). Data are marked by blue points and the total model flux is plotted in gray. The unattenuated light from the old and young stellar populations is shown in magenta and cyan, respectively. The dust component is displayed in red, while the AGN emission in plotted in orange. In the nucleus, the young stellar population is negligible and the AGN produces a large percentage of the near- to mid-infrared flux. On the contrary, in the ring there is a substantial population of young stars and the IR emission is well--fit without any AGN component.
}
\label{fig:SED}
\end{figure}

Regardless of the diagnostic used, the majority of the spectra lie in the LIER region. In particular, the apertures defined inside the ring have the highest [NII]/H$\alpha$ and [OIII]/H$\alpha$ ratios. In this inner region al least part of the ionizing photons could be due to the central AGN \citep[see, e.g.,][]{Bendo2006}. The few apertures with spectra in the star-forming region (shaded in light orange) have been marked with orange circles in Fig.~\ref{fig:apertures}. These apertures correspond to regions with knots of star formation visible in the H$\alpha$ image. In the following figures, apertures in the HII/Comp and LIER regions as determined used the bottom panel of Figure~\ref{fig:BPT} are marked with orange and green symbols, respectively.

\subsection{SED Modeling}
\label{sec:SED}

Spectral energy distribution (SED) fitting was performed using the Code for Investigating Galaxy Emission, or CIGALE \citep{Boquien2019}.  This code assumes energy balance between light emitted in the infrared and absorbed in the UV to produce each model SED. The main properties of bands used to build the SED is detailed in Table~\ref{tab:sedbands}. All of the available photometry described in Section~\ref{sec:phot} was first smoothed to a common resolution of 10\farcs7, the native resolution of the PACS spectroscopy at 158~\micron.  Smoothing was completed using a Gaussian for the FUV--Ks band images and the kernels developed in \citet{Aniano2011} for the IRAC--PACS images.  Fluxes were extracted for each individual region across the disk of M~104. Additional Far-IR photometery is available from the SPIRE instrument on \textit{Herschel} was not included in the SED analysis due to the larger beam of even the shortest--wavelength SPIRE band.  The SPIRE 250~\micron\ data has a PSF of $\sim 18$", much larger than the 10\farcs7 PSF of the PACS 158~\micron\ data.  Using this larger beam would make it impossible to extract fluxes for the individual HII regions seen in the MUSE data. Since the SEDs of the regions across the ring of M~104 peak shortwards of 160~\micron, temperatures and masses of the colder dust should be relatively well constrained without the 250~\micron\ measurements. With this in mind, we choose not to include the 250~\micron\ data and retain the highest possible spatial resolution, even if by using the 250~\micron\ data we could have obtained more accurate estimates of the total far-IR flux.
%To test this decision, we compared the FIR luminosities determined from the results of the SED fitting with and without the inclusion of the 250~\micron\ data.  We found that these two methods produced similar values within 10\%, which is the assumed error for the FIR measurements, and therefore choosing to not include the 250~\micron\ data should not significantly change our results while still allowing for the analysis of structures smaller than the SPIRE resolution.

Due to the presence of an active galactic nucleus (AGN) in the center of M~104, we used two separate parameter ranges to complete the SED fits.  Regions interior to the ring were modeled adding a component of AGN emission, while the regions along the ring were modeled assuming no AGN component. For both region sets, we use the \citet{BruzualCharlot2003} stellar population model with a \citet{Chabrier2003} IMF and the \citet{Draine2014} dust models.  We set the metallicity at $Z = Z_{\odot}$.  The SFH delayed model, which we use for all regions, models the star--formation history (SFH) as beginning at time $t_0$ and peaking at a later time $\tau$, followed by a smooth, exponential decay in the star--formation rate.  We do not include an additional burst in star formation at a later point, nor do we include any quenching.  \texttt{tau\_main} and \texttt{age\_main} set the e-folding time and the age of the main stellar population in this model, respectively.  The dust attenuation is modeled using the \citet{Calzetti2000} attenuation curve.  The dust emission is modeled using \citet{Draine2014} as this model contains detailed information about the PAH emission.  In this model, \texttt{qpah} is the mass fraction of PAHs, \texttt{umin} sets the minimum interstellar radiation field heating the dust, and \texttt{gamma} is the fraction of dust heated by a radiation field greater than \texttt{umin}.  Finally, the AGN component in the inner regions is fit using the \citet{Fritz2006} model where \texttt{f\_AGN} is the AGN fraction and \texttt{psy} is the angle between the equatorial axis of the AGN and the line of sight. The complete set of model parameters for the two regions is listed in Table~\ref{tab:sedprops}.  Two examples of SED fits, one in an inner region (the nucleus) and one along the southern part of the ring (S1) are shown in Figure~\ref{fig:SED}.  All the ETGs were fit using a similar method as the interior regions, with some adjustments to accommodate the limited availability of archival photometry for some of the sources.

%\subsection{\CII{} and dust}
\subsection{\CII{} and UV attenuation}
\label{sec:CIIUV}

\begin{figure}
\begin{center}
\includegraphics[width=0.5\textwidth]{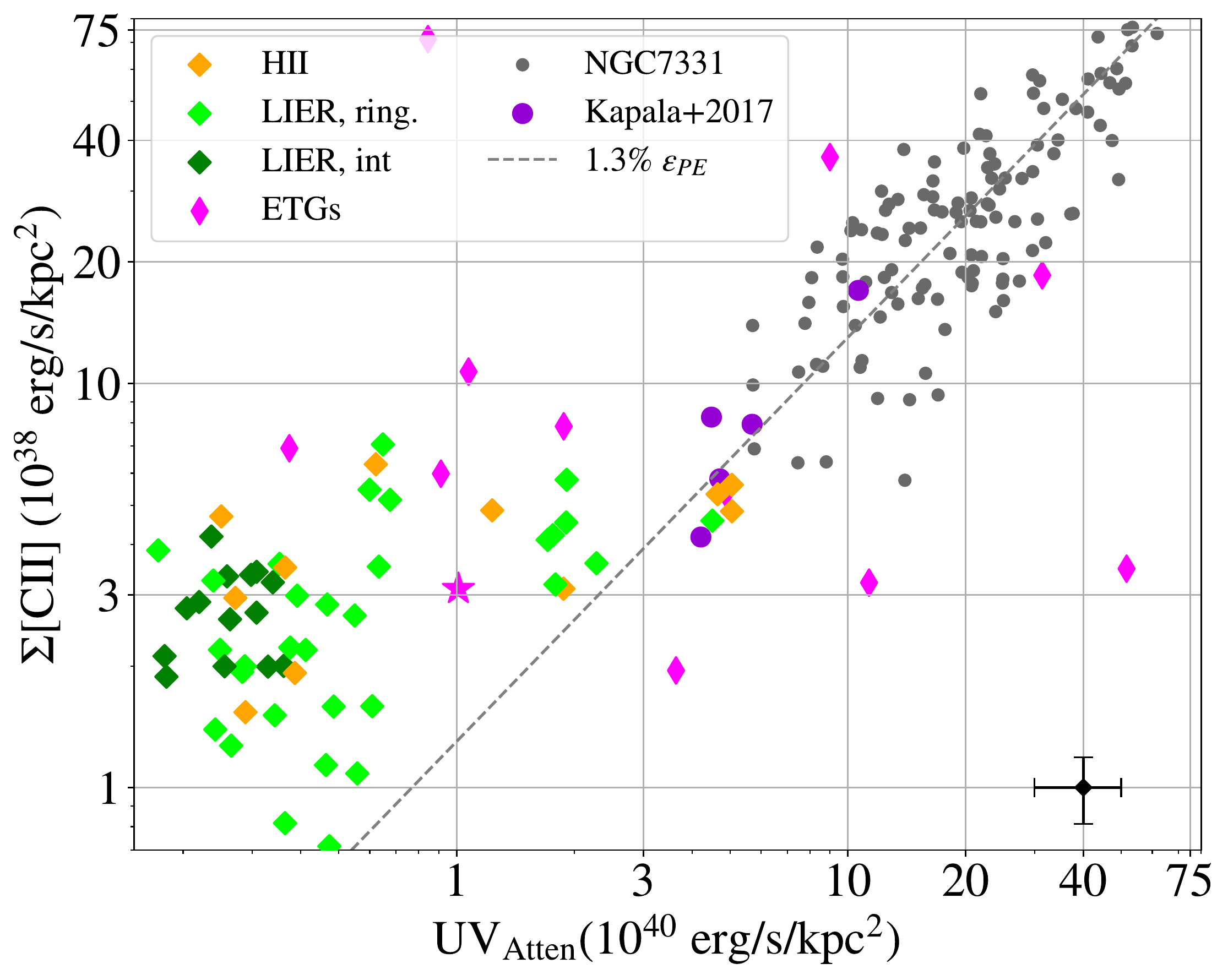}
\end{center}
\caption{Surface brightness of the [CII] line plotted as a function of the attenuated UV light from the CIGALE SED fits. M104 data are color--coded based on BPT diagram diagnostic, with orange being star--forming, and green being LIERs--dominated.  The LIERs are further divided into regions interior to the ring (dark green) and along the ring (light green).  Comparison data from the nearby galaxy NGC~7331, our ETG sample, and M~31 are shown as grey, magenta, and purple symbols, respectively.  The global M104 measurement is shown as a magenta star. Median errorbars, scaled to be the correct size on this plot at the locus of the data from M~104, are shown in the right bottom corner.}
\label{fig:UV_Atten}
\end{figure}
If \CII\ emission is the primary cooling channel for the ISM, it would follow that the strength of the \CII\ emission would be directly related to tracers of heating under the assumption of thermal stability.  This relationship is especially relevant to the application of \CII\ emission as a SFR indicator.  One particularly reliable tracer of gas heating is the amount of attenuated UV light (E~$=6-13.6$~eV), as this high energy light is a major heating source in HII regions and PDRs \citep{Kapala2017}.  The amount of attenuated UV light can be estimated with the CIGALE SED fits, which predict the attenuated light from the modeled populations of young and old stars.  For the regions in our sample, we use the \citet{Calzetti2000} dust attenuation model.  The modeled attenuated light is then integrated from $\lambda = 91.2-206.6$nm to determine the total amount of attenuated UV light.  The results of this analysis are shown in Figure~\ref{fig:UV_Atten}, where the attenuated UV light is plotted against the [CII] surface brightness ($\Sigma$[CII]).  For comparison, the data from resolved regions in NGC~7331~\citep{Sutter2022} and M~31~\citep{Kapala2017} along with our sample of ETGs are shown as gray, purple, and magenta points, respectively.  The global measurement for M~104 is shown as a magenta star, for a direct comparison with the ETG global values.

Comparing these three samples with M~104, we see clear differences.  While the resolved measurements from NGC~7331 and M~31, two star--forming spiral galaxies, fall along a linear relationship with a slope of 0.013, the majority of the regions within M~104 and a subset of the ETGs fall well above this relationship.  The two ETGs that lie in a similar space to NGC~7331 are NGC~1266 and NGC~3032, which have both been suggested to host an AGN \citep{Alatalo2011, Capetti2005}.  In addition, NGC~1266 shows signs of an outflow, which could be producing shock heated gas, increasing the \CII\ emission above that of the other ETGs \citep{Alatalo2011}.  The three ETGs in the right corner of this plot are NGC~4636, 5322, and 5576.  It is less clear why this subsample of ETGs fall in this corner, but this could be due to differences in modeling when considering resolved and global galaxy measurements.  The differences between M~104 and the remaining ETGs and the resolved measurements from NGC~7331 and M~31 suggests that the primary heating source differs between M~104 and other early--type galaxies and star--forming galaxies.  While young O and B stars provide the main heat source in NGC~7331 and M~31, M~104 has very little ongoing star formation leading to the dominant source of heat being evolved stars.  This would explain the lower values of attenuated UV light across M~104, as there is little UV light to be attenuated to begin with.  

\subsection{[CII] and total infrared luminosity}
\label{sec:CIIFIR}

\begin{figure*}
\begin{center}
\includegraphics[width=0.8\textwidth]{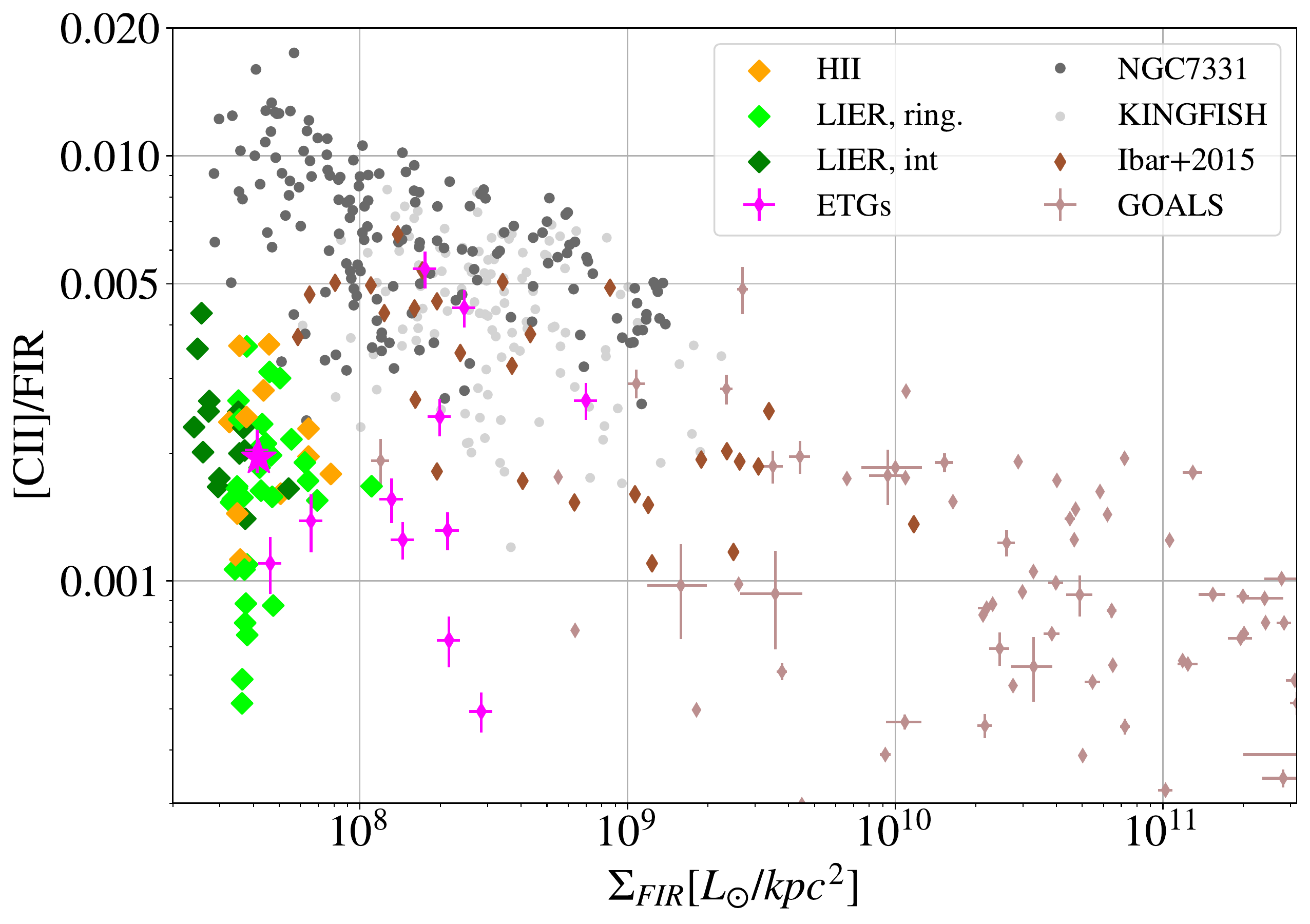}
\end{center}
\caption{The [CII]/FIR ratio plotted as a function of far-IR surface brightness ($\Sigma_{FIR}$).  The regions within M~104 are color--coded based on their position in the BPT diagram as in previous figures. 
The comparison data from NGC~7331~\citep{Sutter2022}, KINGFISH (star--forming regions in local galaxies from \citet{Sutter2019}), $z\sim0.02 - 0.2$ galaxies from \citet{Ibar2015}, and local U/LIRGS \citep{DiazSantos2017} are shown in gray, light gray, brown, and light brown, respectively. We also include global measurements from local early type galaxies (ETGs) in magenta, and the global measurement of M104 as a magenta star.
}
\label{fig:CIIdef}
\end{figure*}

In order to explore the relationship between \CII\ emission and dust heating, we show the \CII/FIR plotted against the FIR surface density in Figure~\ref{fig:CIIdef}.  The FIR fluxes were computed by integrating the modelled SEDs from $\lambda=8-1000 \mu$m.  The FIR surface brightness was then determined by dividing the FIR luminosity by the deprojected area of one region.  To compare the regions within the M~104 to previous studies of the \CII/FIR relationship, we also plot the data from the resolved regions across the disk of nearby star--forming galaxy NGC~7331 \citep[dark gray points,][]{Sutter2022}, resolved star--forming regions from the ``Key Insights in Nearby Galaxies: a Far-Infrared Survey with \textit{Herschel}'' \citep[KINGFISH, light gray points,][]{Sutter2019}, global measurements of $z\sim0.02-0.2$ galaxies \citep[brown diamonds,][]{Ibar2015}, and global measurements from local luminous infrared galaxies (LIRGS) from the Great Observatory All--Sky LIRG Survey \citep[GOALS, tan diamonds][]{DiazSantos2017}.  In order to perform a uniform analysis of these diverse data sets, we deprojected the infrared surface brightness measurements reported in \citet{Ibar2015} and \citet{DiazSantos2017} by multiplying by a factor of $1 / \cos{i}$, where $i$ is the galaxy's inclination.  For the sources included in \citet{Ibar2015}, the inclinations were determined by fitting ellipses to the PANSTARRs $r$ band images of each galaxy and assuming $\cos{i} = b/a$.  For the galaxies in the GOALS sample, the inclinations were taken from \citet{Kim2013}.  With these updated $\Sigma_{FIR}$ measurements, we see a linear trend between \CII/FIR and $\Sigma_{FIR}$ across three orders of magnitude in $\Sigma_{FIR}$ for our comparison samples.  

\begin{figure}[t!]
\begin{center}
\includegraphics[width=0.49\textwidth]{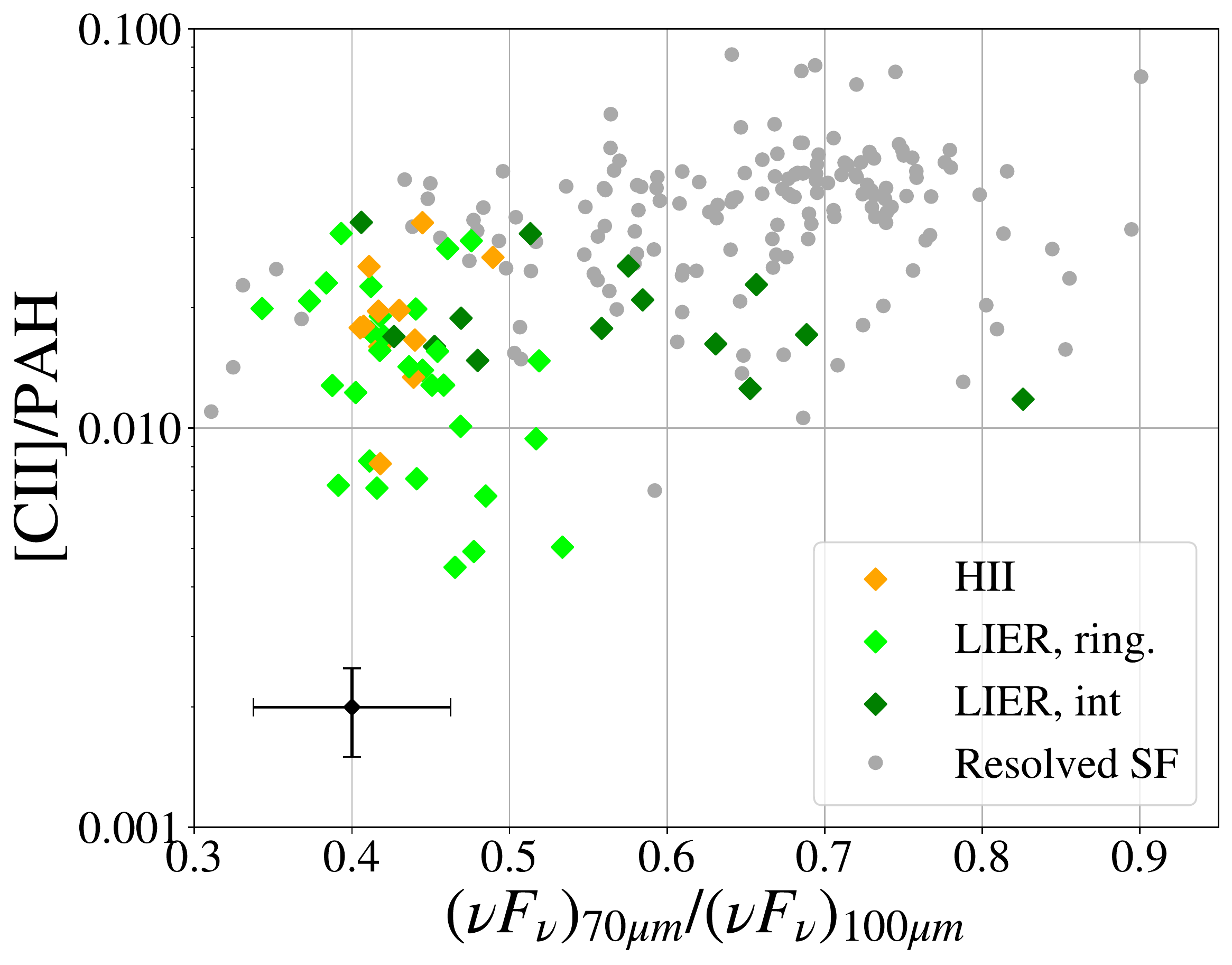}
\end{center}
\caption{
The [CII]/PAH values plotted as a function of $(\nu F_{\nu})_{70\mu m} / (\nu F_{\nu})_{100\mu m}$.  The points from M~104 follow the color scheme of previous plots. Resolved measurements from three nearby star--forming galaxies, NGC~7331, NGC~4559, and NGC~1097 \citep{Sutter2022, Croxall2012} are shown as gray points. Median errorbars for M104 are shown on the left.
}
\label{fig:CIIpah}
\end{figure}

By comparing our measurements of \CII/FIR across M~104 to the trends measured in other nearby galaxies, it becomes clear that M~104 is an outlier with significantly lower values of \CII/FIR for a given $\Sigma_{FIR}$.  In order to determine what unique properties of M~104 might be causing it to fill this otherwise empty area of this plot, we consider similar behavior observed in our ETG sample (see Section~\ref{sec:etgs} for more information about this sample).  The global \CII/FIR and $\Sigma_{FIR}$ points from this sample are displayed as magenta diamonds in Figure~\ref{fig:CIIdef}, while the global measurement for M~104 is shown as a magenta star for a consistent comparison.  A subset of the ETG sample does fall in a similar space as the data from M~104, suggesting that the cause of the observed low values of \CII/FIR in M~104 is shared in other ETGs.  The two ETGs that fall among the gray points indicating star--forming galaxies are NGC~1266 which has an outflow and hosts an AGN \citep{Alatalo2011}, NGC~3032 which has some observable spiral structure as well as an AGN \citep{Capetti2005}, and NGC~4710, a nearly edge on galaxy which makes measurements of $\Sigma_{FIR}$ more uncertain.  This could suggest that other mechanisms are exciting the \CII\ emission in these three galaxies, making them outliers from the rest of the ETG sample.

With the additional information from the MUSE data, we can also subdivide the regions in M~104 based on what the dominant source of ionizing radiation is (see Section~\ref{sec:BPT} for details).  In this figure, the regions in M~104 classified as HII or composite are orange and regions that fell in the LIER category are green.  We further subdivide the LIER category into regions along the ring (light green diamonds) and regions interior to the ring (dark green diamonds), with the nucleus marked with a dark green star.  Using these diagnostics, we see that the regions dominated by star formation have on average the highest \CII/FIR values, while those classified as LIERs have the lowest.  Studies of resolved BPT diagnostics have shown that regions classified as LIERs are dominated by emission from an older stellar population, predominantly AGB stars \citep{Belfiore2016, Zhang2017}.  This could drive the \CII/FIR measurements lower as AGB stars will heat the dust, producing FIR emission, while emitting very little radiation with energy high enough to ionize carbon, limiting the production of the \CII\ line.  Within the LIER sample, the regions in the interior show slightly higher \CII/FIR values than those in the ring, which could suggest some additional \CII\ emission excited by the central AGN.

\begin{figure}
\begin{center}
\includegraphics[width=0.49\textwidth]{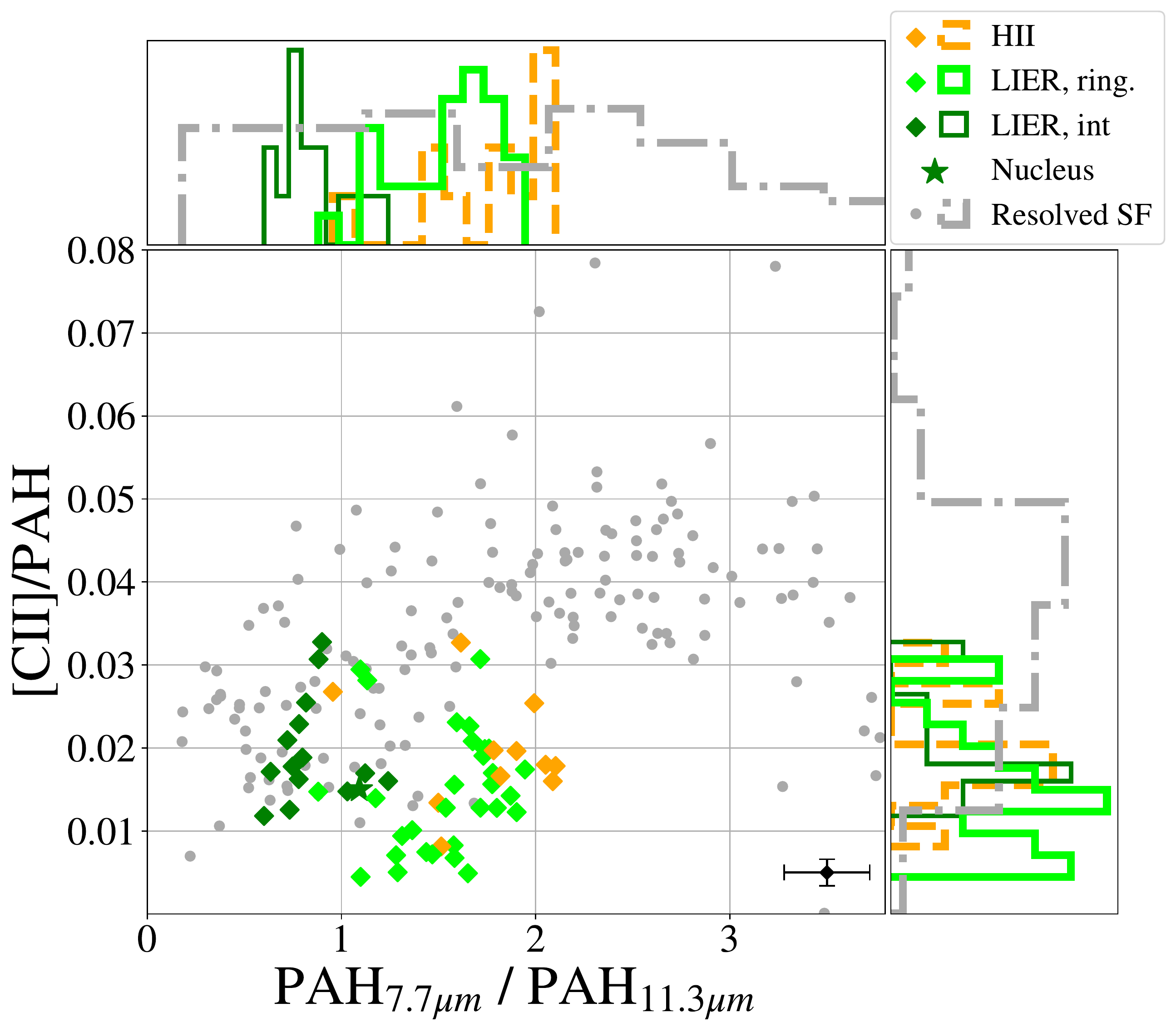}
\end{center}
\caption{The [CII]/PAH values, now plotted against the ratio of the 7.7~\micron\ and 11.3\micron\ PAH features, an indicator of PAH charge.  Same color scheme as previous plot.  Histograms above and to the right show the distribution of the different types of regions in M~104 with respect to the star--forming galaxy sample.  Median errorbars for the data from M104 are shown in the lower-right corner.}
\label{fig:PAH_charge}
\end{figure}

\subsection{[CII] and PAH emission}
Studies have suggested that comparing \CII\ emission to the emission from polycyclic aromatic hydrocarbons (PAHs) might provide an improved indicator of the photoelectric heating efficiency in PDRs \citep{Croxall2012}.  As PAHs provide the bulk of free electrons in PDRs, the emission from these small dust grains provide a key indicator of the conditions within PDRs where much of the \CII\ emission is produced \citep{Sutter2019, Tarantino2021}.  By comparing the \CII\ and PAH observations across M~104 we can therefore determine how the PDR conditions might be affecting the strength of the observed \CII\ emission.

As the 7.7~\micron\ and 11.3~\micron\ PAH features are nicely covered by the IRAC4 and WISE3 bands, respectively, we use the fluxes from these bands along with the results of the SED fits to estimate the PAH fluxes. To remove contributions from stars, large dust grains, and AGN emission from these bands, the modelled flux from each of these components is summed, convolved with the transmission function for the specified band, and then subtracted from the observed flux. The remaining emission is then assumed to be only the emission from the 7.7 and 11.3~\micron\ PAH emission features, which are added together to estimate the total PAH emission.  The results of this analysis are used to compare the \CII\ and PAH emission across M~104.

Figure~\ref{fig:CIIpah} shows the \CII/PAH values plotted as a function of $(\nu F_{\nu})_{70\mu m} / (\nu F_{\nu})_{100\mu m}$, an indicator of dust temperature.  For comparison, resolved regions from star--forming galaxies NGC~1097 and NGC~4559~\citep{Croxall2012} as well as NGC~7331~\citep{Sutter2022} are shown as gray points.  The ETG sample is not included in these measurements due to uncertainties in our PAH estimation when considering global measurements.  Specifically, the majority of the ETGs have sparse coverage in the mid-- to far--infrared photometric bands, making the precise determination of the fraction of the flux from PAHs in the available IRAC4 and WISE3 fluxes prone to significant errors.

Unlike the \CII/FIR measurements shown in Figure~\ref{fig:CIIdef}, we see that there is some overlap between the regions in M~104 and regions in the star--forming galaxies.  This overlap includes all of interior regions classified as LIERs and the majority of the regions classified as HII--dominated using the BPT classification mechanism.  The similarity between M~104 and star--forming galaxies suggests that changes to the PAHs are not the driving mechanism for the observed differences in \CII/FIR, and that the ISM conditions that suppress \CII\ emission are likely also limiting the PAH emission; either through the destruction of small grains near an AGN \citep{Langer2015} or a lack of photons with energies high enough to excite the PAHs \citep{Malhotra2001, Smith2017}.

In addition to comparing the \CII/PAH emission to the dust temperature, we also compare to the ratio of the 7.7\micron\ and 11.3\micron\ PAH features, an indicator of average PAH charge \citep{Draine2021}.  This is shown in Figure~\ref{fig:PAH_charge}.  Again, we see overlap between the regions from M~104 and the comparison sample of resolved star--forming galaxies.  This is especially true for regions classified as LIERS inside the ring.  One major difference between the data from M~104 and the star--forming sample is the range of 7.7/11.3 we measure.  While the regions from the three star--forming galaxies are fairly evenly distributed from $\sim 0-4$, regions from M~104 all fall between 0.5 and 2.2, implying a much more uniform population of PAHs across the disk.  The range in charges is even more constrained when we consider only regions along the ring (light green and orange points) or only regions interior to the ring (dark green points), suggesting that the population of PAHs is remarkably consistent within these two environments.  The lack of 7.7/11.3 values above 2.5 shows that the PAHs within M~104 have a lower charge on average, and therefore PAH grain charging is likely not the cause of the low \CII/FIR values observed in Figure~\ref{fig:CIIdef}.

\subsection{Measuring the \texorpdfstring{$H_2$}{} Mass}
\label{sec:H2Mass}
We can estimate the total mass of molecular hydrogen in the ring by integrating the total CO emission. We assume a factor 2 correction to account for the likelihood that the $^{12}$CO$_{1\rightarrow 0}$ line will become optically thick in dense star--forming regions~\citep{Hughes2017}. By applying the conversion from Eq.~3 from \citet{Bolatto2013}, we find out that the total molecular hydrogen mass of the ring of M~104 is $0.9\times10^9$~M$_\odot$. This places M~104 on the high end of molecular hydrogen mass when compared to the early--type galaxies from the ATLAS$^{3D}$ catalog~\citep{Lagos2014}. On the other hand, the mass of molecular hydrogen in the 10~kpc ring of M~104 is approximately half of the total mass of molecular hydrogen found in the 5~kpc ring of our own galaxy~\citep[$\approx 2\times10^9$~M$_\odot$,][]{Clemens1988}. So, although M~104 cannot be classified as a star-forming galaxy, there is a substantial amount of molecular hydrogen contained in the ring of M~104.

\begin{figure}[t!]
\begin{center}
\includegraphics[width=0.49\textwidth]{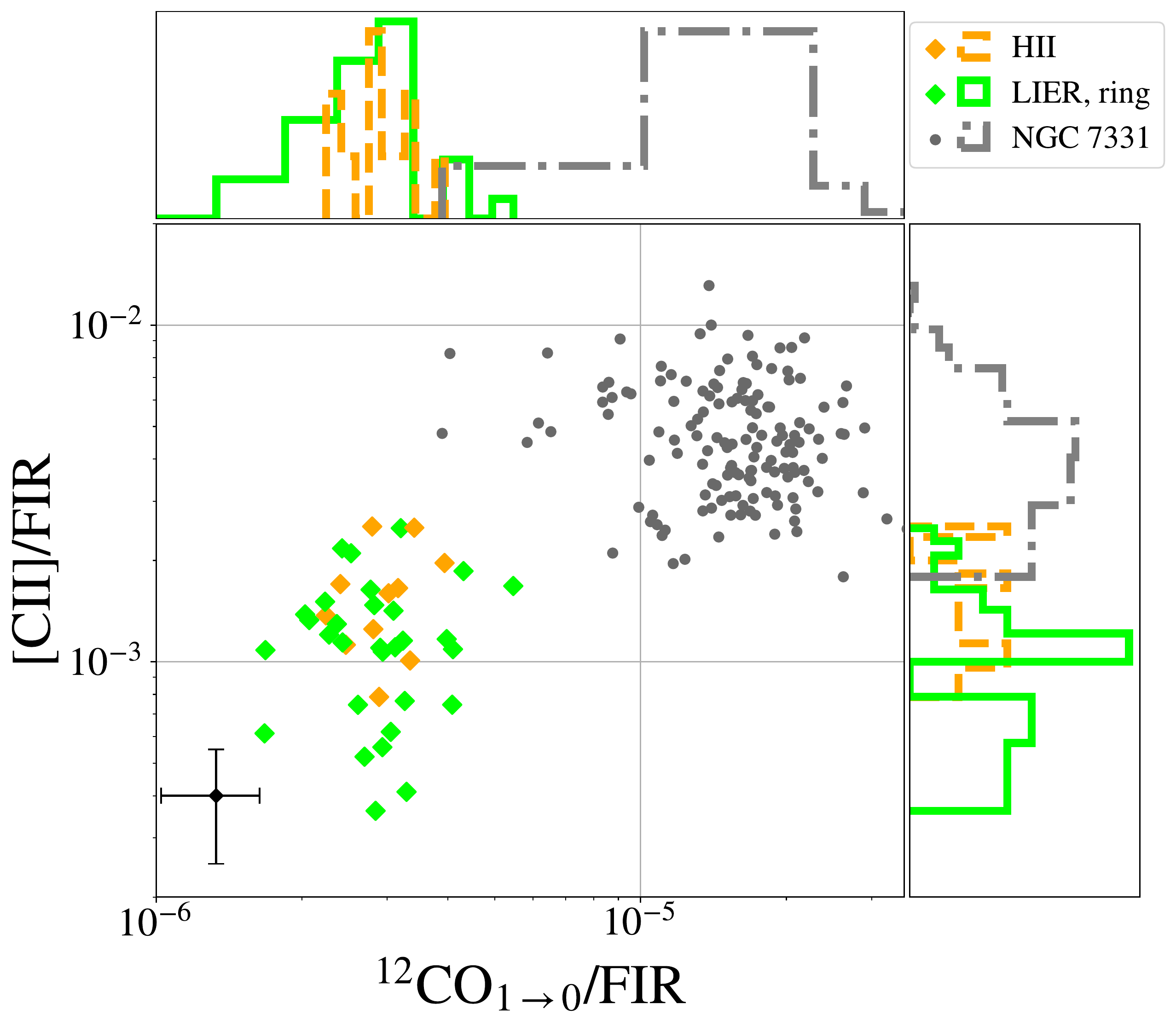}
\end{center}
\caption{CO luminosities plotted against the [CII] luminosities, both normalized by the FIR luminosity. Median errorbars for the data from M104 are shown in the lower left corner.
}
\label{fig:CII_CO}
\end{figure}

\subsection{\CII{} and CO emission}

As \CII\ emission can originate in a variety of environments including molecular gas, we compare the \CII\ and $^{12}$CO$_{J=1\rightarrow0}$ (CO, hereafter) in regions where CO was detected.  These regions primarily consist of regions along the ring as well as one region near the nucleus with a weak CO detection (region R1, see Figures~\ref{fig:maps} and \ref{fig:apertures}).  As the CO flux in region R1 is much lower than in the ring and approaching the detection limit of ALMA, we do not include it in our analysis.  This leaves only regions along the ring with significant \CII\ and CO detections, so the following analysis will not include the interior regions discussed in previous sections.  It is apparent by considering the \CII\ and CO maps that the distribution of \CII\ and CO are very different in M~104, suggesting that the majority of the \CII\ emission is not tracing the molecular gas phase of the ISM in this galaxy.  

In order to further compare the \CII\ and CO emission, we plot the the \CII\ against the CO emission, both normalized by FIR luminosity, in Figure~\ref{fig:CII_CO}. We choose to plot the normalized fluxes instead of comparing the \CII\ and CO luminosities because the comparison between \CII/FIR and CO/FIR is often used as an input in PDR models, making this plot a good comparison for other studies \citep[see e.g.][]{Madden2020}.  In addition, as CO is only observed around the ring and has a relatively constant flux around ring, the direct comparison of the two lines is dominated by scatter and there is little observed relation between the two line luminosities.  As in previous figures, data from NGC~7331 is shown as a comparison sample.  By examining the distribution of the points from NGC~7331 and M~104, it is clear M~104 is proportionally much lower in both \CII\ and CO emission.  It is also notable that while the CO/FIR values from NGC~7331 are distributed over a range of $3.9\times10^{-6} - 6.7\times10^{-5}$, M~104 only covers a range of $1.4\times10^{-6} - 5.6\times10^{-6}$, which is both lower and much more constrained than NGC~7331.  This is also apparent in the histogram plotted above Figure~\ref{fig:CII_CO}, where we can see the small range of the data from M~104, and that there is little overlap with the data from NGC~7331.  This is primarily due to the relatively constant CO fluxes across the ring (see the top panel of Figure~\ref{fig:maps}).  

Although the \CII/FIR values in M~104 are also lower than those in NGC~7331, there is some overlap between the two galaxies, especially in regions in M~104 that are classified as star--forming using the emission line diagnostics.  This difference between the \CII\ and CO behavior suggests that while the \CII\ emission is tracing the small amount of star--formation still occurring in M~104, the molecular gas as traced by CO emission is not.  For this reason, we choose to not include PDR models on this plot, as it seems the CO emission may not be tracing this phase of the ISM.
\begin{figure}[t!]
\begin{center}
\includegraphics[width=0.49\textwidth]{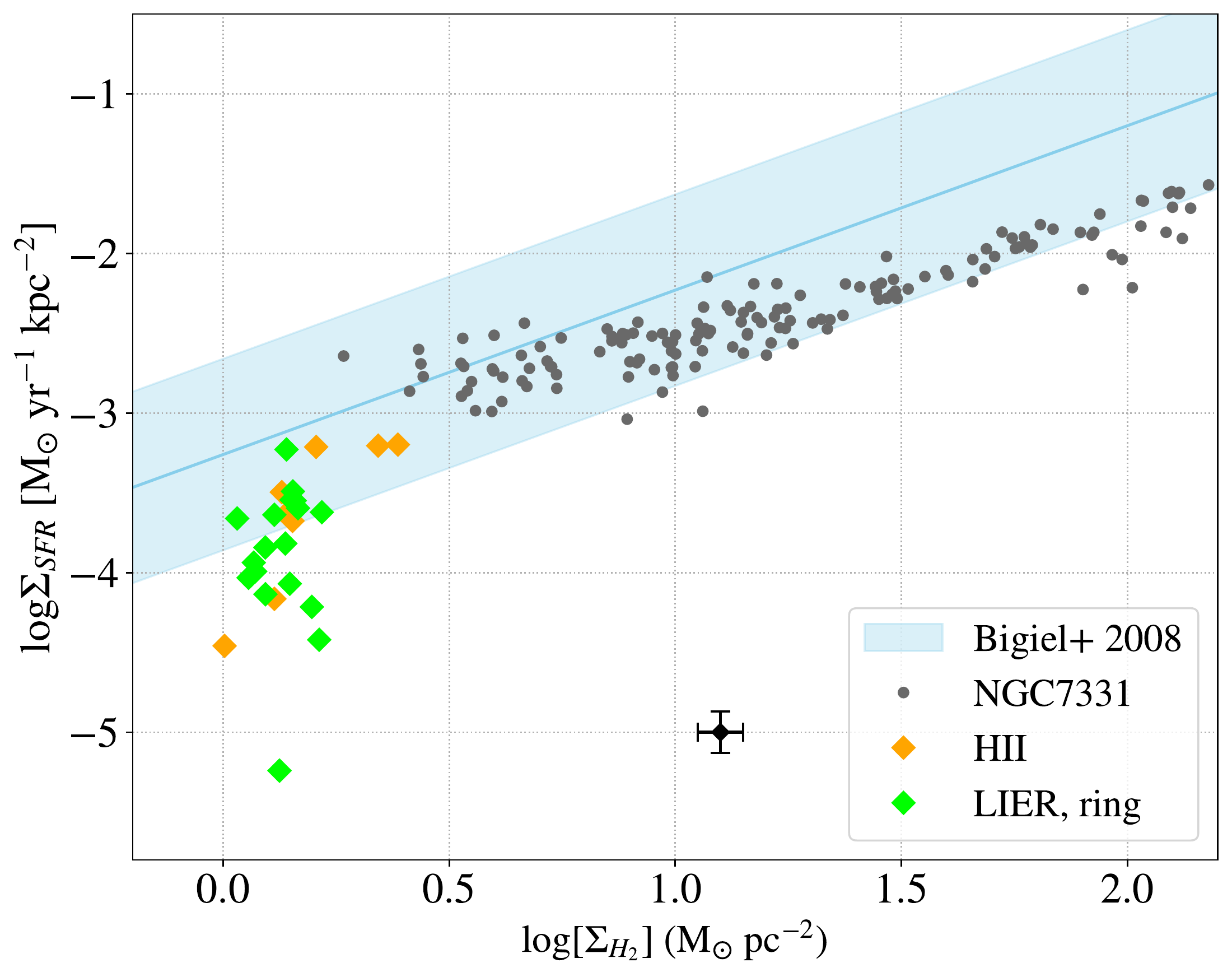}
\end{center}
\caption{A Kennicutt-Schmidt plot with star--formation rate surface density plotted against the molecular gas surface density. The relationship for sub-Kpc regions in starforming galaxies from \citet{Bigiel2008} is shown as a blue line, with blue shading representing the 3$\sigma$ error on the slope. Data for NGC~7331 \citep{Sutter2022} are marked with gray dots. Median errorbars for the data from M104 (orange and green diamonds) are shown as a black cross in the lower part of the plot.}
\label{fig:KS}
\end{figure}

To further test the disconnect between CO and star formation in M~104, we plot the surface density of star formation vs the surface density of molecular gas in Figure~\ref{fig:KS}.  This plot is commonly referred to as a Kennicutt--Schmidt plot, and the proportional trend between $\Sigma_{\rm{SFR}}$ and $\Sigma_{H_2}$ demonstrates how gas in the ISM is the fuel for ongoing star formation \citep{Kennicutt2012}.  

For SFR values we used the CIGALE estimates which take into account the entire SED from the far-IR to the far-UV. The H$_2$ gas mass is determined by measuring the molecular gas mass for each region using the method described in Section~\ref{sec:H2Mass}.  Both the H$_2$ gas mass and the SFR are converted to surface densities by dividing by the deprojected area of the aperture used in the measurement. For comparison, we plot the relationship from \citet{Bigiel2008} which investigated the Kennicutt-Schmidt relationship at sub-Kpc scale for several star-forming galaxies as a blue line, with the dispersion represented using blue shading, as well as the relationship for regions of the Milky Way analog NGC~7331 \citep{Sutter2022}.
Figure~\ref{fig:KS} clearly shows that there is no correlation between $\Sigma_{H_2}$ and $\Sigma_{\rm{SFR}}$ in the regions within M~104. This supports the conclusion that the majority of the CO emission is not tracing the ongoing star formation, and therefore is not a useful indicator of PDR conditions.

\section{Summary and Conclusions}
\label{sec:conclusions}

We presented the analysis of unpublished archival observations of the galaxy Messier~104, also known as the Sombrero galaxy, obtained with ALMA, {\it Herschel}, and MUSE. These observations have been compared to a wealth of archival photometric observations from the UV to the far-IR. The measurements used in the Figures are available in Appendix~\ref{sec:measurements} (see Tables~\ref{tab:sample}). The main results of our study can be summarized as follows:
\begin{itemize}
    \item The emission of dust and molecular gas is very uniform all along the ring. In particular, the ring is clearly defined from 8~$\mu$m to the far-IR. The CO emission from molecular gas is mainly confined to the ring with the exception of a weak emission from a circumnuclear region, while dust emission is also visible in the nucleus and in the region inside the ring. The \CII{} and H$\alpha$ emission are visible all around the ring but they appear less uniform, with intense spots linked to knots of star formation. 
    \item The kinematics of stars and gas along the ring is extremely regular. The ionized gas traced by the H$\alpha$ emission as well as the gas traced by CO and \CII{} have the same rotational velocity inside the measurement errors. The velocity dispersion is constant along the ring, while the inner part of the disk has a larger velocity dispersion.
    \item Through comparisons with a set of early--type galaxies observed with {\it Herschel} and SOFIA, we find that both M~104 and a sub--sample of the early--type galaxies populate a region of the \CII/FIR vs $\Sigma_{FIR}$ under the locus of star-forming galaxies. 
    This is likely due to the fact that although the old stellar population heats the dust which emits continuum far-IR radiation, it is not able to excite the \CII\ line, leading to the lack of correlation between \CII{} and FIR emission and therefore to low \CII/FIR values.
    \item  The study of the \CII{} emission to dust emission and gas heating indicators such as UV attenuation, far-IR total emission, and PAH emission, show that M~104 behaves like early-type galaxies. However, the inner region and the parts of the ring with visible star formation in H$\alpha$ have higher \CII/FIR and \CII/PAH ratios, closer to those of star-forming galaxies.
    \item The total mass of molecular hydrogen in the ring of M~104 is $0.9\times10^9$~M$_\odot$, a value intermediate between typical early-type galaxies and the molecular hydrogen content of the ring of our own galaxy.
    \item  While the \CII{} emission is comparable to those of faint star-forming regions in other galaxies, the CO emission is much lower. The mass of H$_2$ derived from the CO emission does not correlate with the star formation along the ring. It is therefore clear that the molecular hydrogen of the ring is not converted efficiently into stars and that most of the \CII{} emission is probably not related to molecular gas, but rather to ionized and neutral atomic gas. 
\end{itemize}

The study of this iconic galaxy shows how a direct interpretation of CO and \CII{} emission should be used with caution in the case of galaxies with low star formation. The lack of correlation between CO emission and star formation as well as the fact that the \CII{} emission is probably not linked to the reservoir of molecular hydrogen should be taken into account when studying the evolution of early-type galaxies in high redshift surveys.

\begin{acknowledgments}
The authors are grateful to Dr. Yuguang Chen for his help in interpreting the MUSE data and to the anonymous referee for the numerous and detailed suggestions which greatly improved the quality of the paper. This research is based on data and software from: the SOFIA Observatory, jointly operated by USRA (under NASA contract NNA17BF53C) and DSI (under DLR contract 50 OK 0901 to the Stuttgart University); Herschel, an ESA space observatory with instruments provided by European-led P.I. consortia and important NASA participation; the Spitzer Space Telescope, operated by JPL/Caltech under a contract with NASA; 
WISE, a joint project of UCLA and JPL/Caltech, funded by NASA; the SDSS survey, funded by the Sloan Foundation,  participating institutions, NSF, DOE, NASA, Monbukagakusho (Japan), Max Planck Soc. (Germany), and HEFCE (UK); 2MASS, a joint project of the Univ. of Massachusetts and IPAC/Caltech, funded by NASA and NSF; GALEX, a NASA small explorer, whose archive is hosted by  HEASARC; MUSE observations collected at the ESO observatories under program 60.A-9303(A); data ADS/JAO.ALMA\#~2018.1.00034.S from ALMA, a partnership of ESO, NSF (USA), NINS (Japan), NRC (Canada), MOST and ASIAA (Taiwan), and KASI (Rep. of Korea), in cooperation with the Rep. of Chile. The Joint ALMA Observatory is operated by ESO, AUI/NRAO and NAOJ. NRAO is a NSF facility operated under cooperative agreement by Associated Universities, Inc. Partial financial support for this project was provided by NASA through the award \#~SOF-06-0032 issued by USRA.
\end{acknowledgments}


\vspace{5mm}
\facilities{SOFIA (FIFI-LS), Herschel (PACS, SPIRE), Spitzer (MIPS, IRAC), WISE, ALMA, GALEX, SDSS, 2MASS, VLT(MUSE)}


\software{
astropy \citep{astropy2013,astropy2018}, sospex \citep[\url{www.github.com/darioflute/sospex,}][]{Fadda2018}, HIPE (\url{www.cosmos.esa.int/web/herschel/hipe-download}), CIGALE \citep[\url{https://cigale.lam.fr/,}][]{Boquien2019}
}


 
\appendix
\restartappendixnumbering
\section{Transient correction of PACS data}
\label{sec:transients}
\begin{figure*}
\begin{center}
\includegraphics[width=0.95\textwidth]{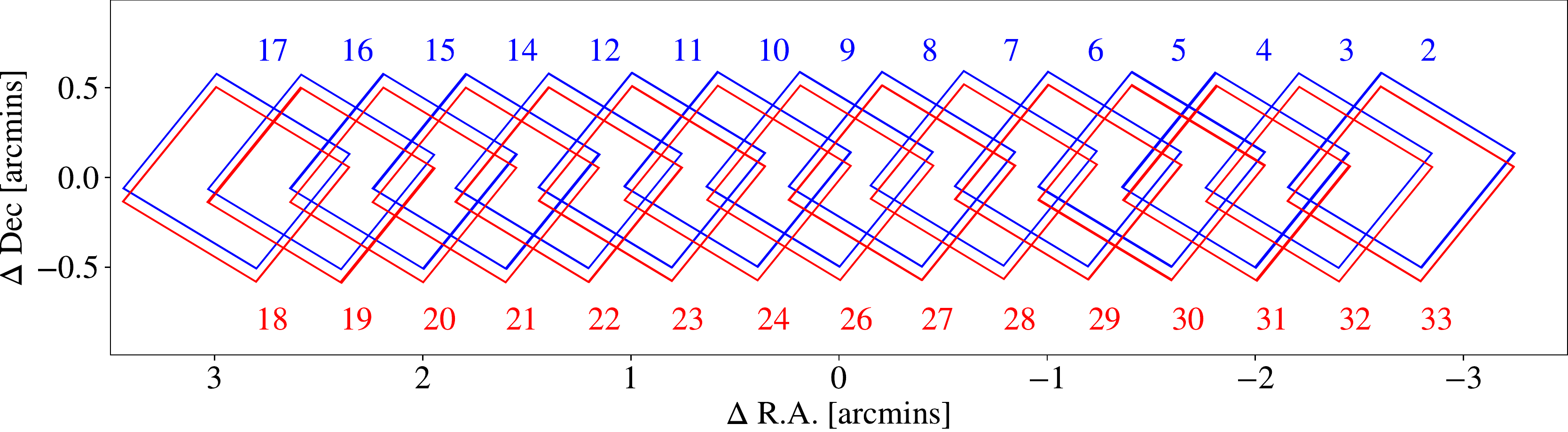}
\includegraphics[width=0.95\textwidth]{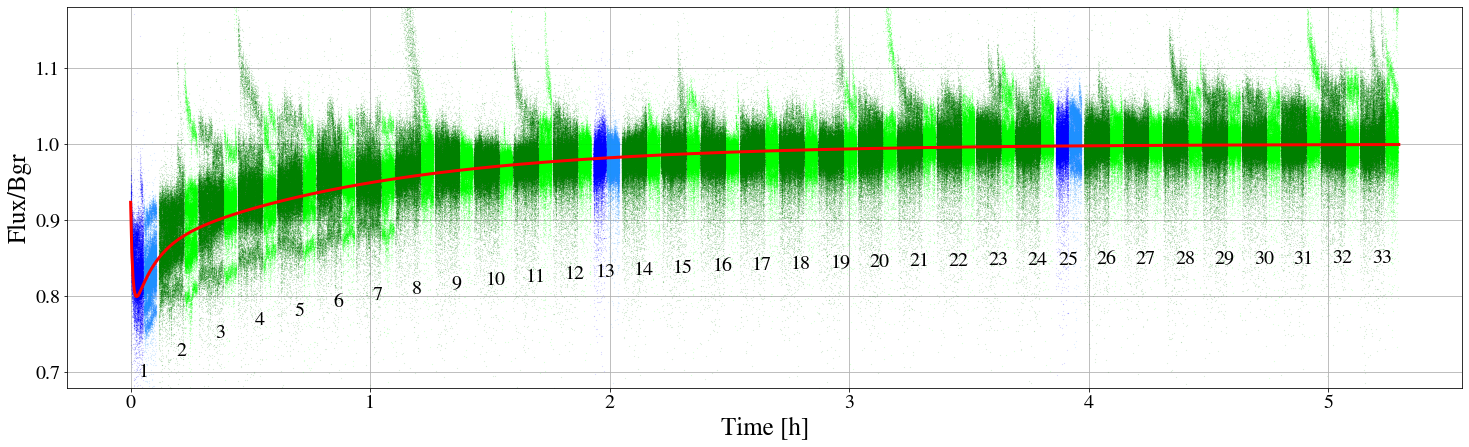}
\end{center}
\caption{Scan technique used to observe M~104 with PACS. Each square is a position of the array on the sky with respect the center of the galaxy. The scan has been repeated twice in opposite directions (blue and red squares). The numbers refer to the number of the slice used in the bottom panel figure which shows the signal normalized to the median signal for all the 16 spexels in spaxel 2. The two different tonalities of green and blue indicate the two wavelength ranges observed in each position of the scan. In particular, the green slices are on-target positions while the blue ones are off-target positions (scan blocks 1, 13, and 25).  The red line shows the fit of the transient behavior of the response. Note that the same position observed at the beginning (2) and end (33) of the observation has a variation of 15\% in the response of the detectors, while the flux in the off-target positions varies by 20\%.}
\label{fig:scan}
\end{figure*}

In this appendix we describe in detail the process used to reduce the \CII{} data obtained with PACS on Herschel. The data have been obtained by scanning the rectangular region containing the ring twice. At the second scan, the center was moved along the perpendicular direction to add a minimum spatial dithering (see Figure~\ref{fig:scan}).
At each position, two spectral regions were observed. During the observation the primary wavelength were 157.74~$\mu$m (\CII{} line) and 88~$\mu$m ([OIII] line). Since PACS had a dual array, When observing the [OIII] line in the blue array, the parallel observation in the red array was centered at 177.4$\mu$m. In the following we refer to observations in one wavelength range and in one raster position as science blocks.
In Figure~\ref{fig:scan} each science block relative to the 158$\mu$m observation is colored in light green, while the parallel observation is colored in dark green. The off-target positions for the two wavelength regions are drawn in light and dark blue, respectively.

During the observation, the response of the detectors varies because of sudden variations in the incident flux. This mainly happens in three situations.

The first reason is linked to the change of the observed field and wavelength range at the beginning of the observation. During the slewing from the previous target, calibration observations of the internal black-bodies are taken. Then, an initial off-target position is observed before targeting the source. In this specific observation, the telescope was pointed to an off-target position two other times.
A long-term transient is visible along the entire observation and it is due to the sudden variation of flux when the detector moved from the previous observation (see bottom panel of Fig.~\ref{fig:scan}).

\begin{figure}
\begin{center}
\includegraphics[width=0.6\textwidth]{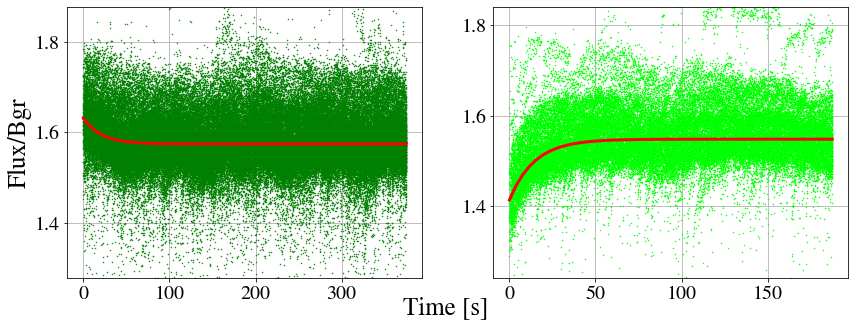}
\includegraphics[width=0.6\textwidth]{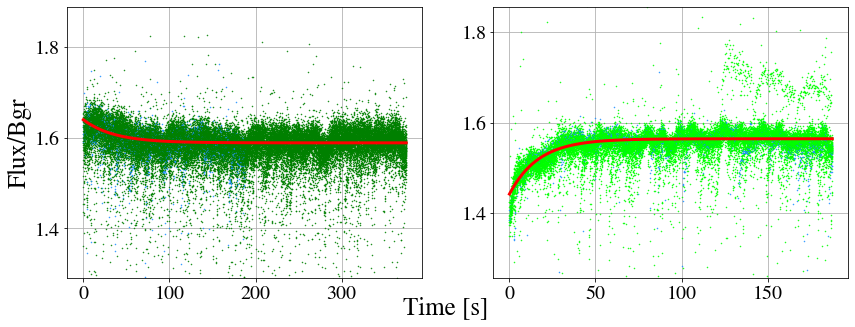}
\end{center}
\caption{The transient at the beginning of each science block The color code is as in Figure~\ref{fig:scan}. On the top, the first estimates of the two science-block transients are obtained by fitting the last science blocks of all the 16 spexels of spaxel 2. On the bottom, the model is refined for the single spexel 6 of spaxel 2 using most of the science blocks of the observation corrected for the long-term transient.
}
\label{fig:slice}
\end{figure}

The second reason is due to the alternate observation of two different wavelength ranges. When observing the region centered at 157.7$\mu$m, for each raster position the grating is moved to also cover the wavelength region centered at 177.4$\mu$m. Since the two regions have quite different telescope backgrounds, the simple fact of alternating between the two bands introduced a transient behavior in the response of the detector. As visible in Fig.~\ref{fig:slice}, the signal at the beginning of each science block can have a 10\% different response with respect to that at the end of the block.

\begin{figure*}
\begin{center}
\includegraphics[width=0.91\textwidth]{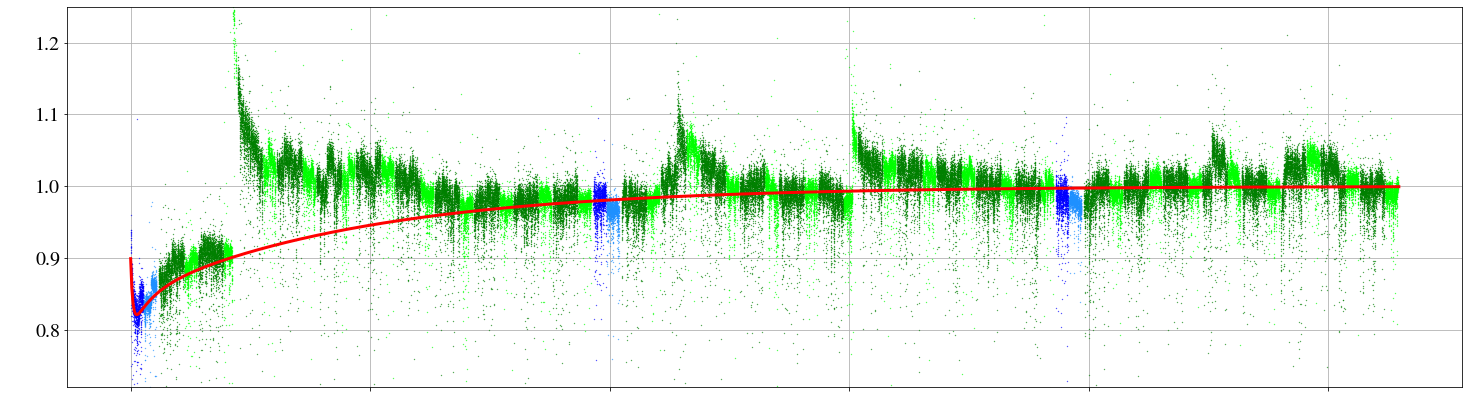}
\includegraphics[width=0.91\textwidth]{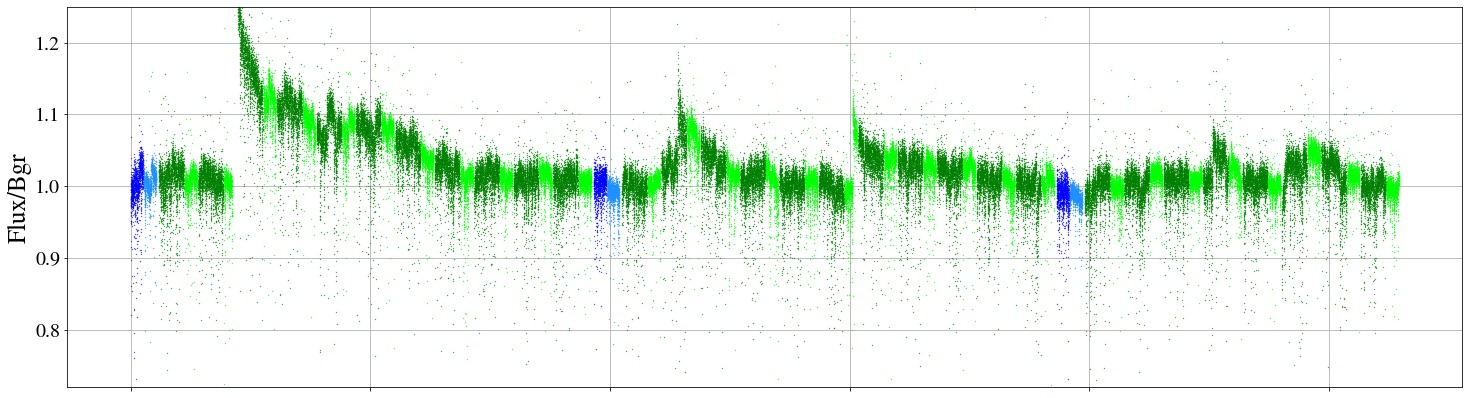}
\includegraphics[width=0.91\textwidth]{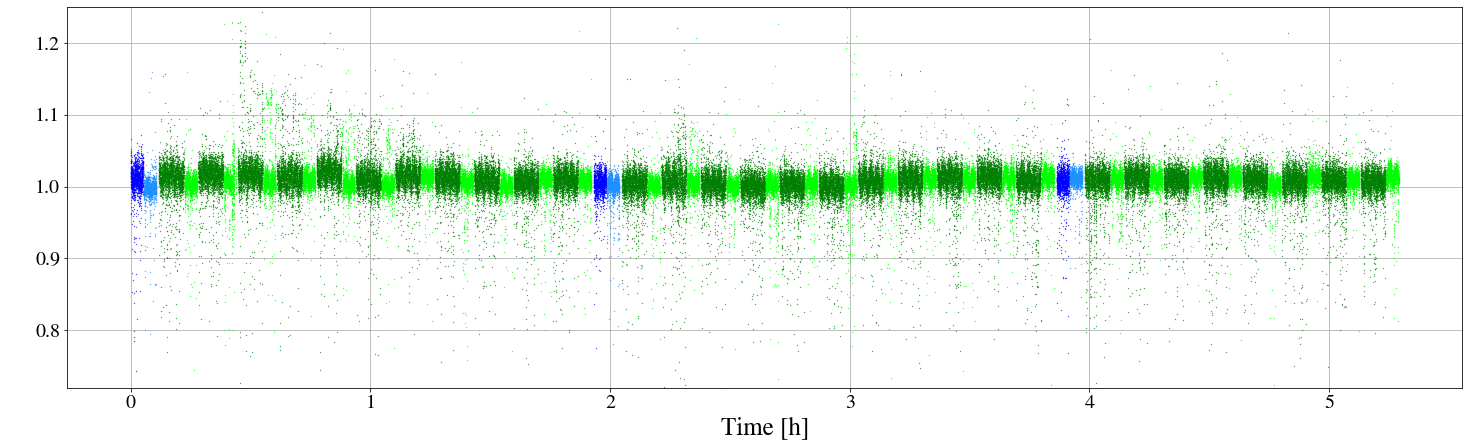}
\end{center}
\caption{Effect of transient corrections on a single pixel: spaxel 2, spexel 6. The original signal normalized to the telescope background (top panel) is fitted with a transient model (red line). The correction (middle panel) puts in evidence the transients caused by cosmic rays hitting the detector. These are corrected using the HIPE pipeline leaving only the flux variations along the scan at the end of the process (bottom panel).}
\label{fig:pixel}
\end{figure*}

Finally, cosmic rays hit the detector randomly throughout the observations, creating a sudden surcharge which dissipates slowly (typically in less than an hour, see middle panel of Fig.~\ref{fig:pixel}). Such transients can be fitted with a combination of exponential functions as explained in \citet{Fadda2016}. 

We will therefore differentiate between a long-term transient affecting the entire observation, science-block transients affecting the beginning of each science block, and cosmic ray transients produced by the random impact of cosmic rays on the detector.
To disentangle these different effects one can use the statistics of repeated observations.
Moreover, since the flux of the observed target is very weak and the background of the telescope dominates total flux, the total flux does not differ too much from the off-target positions for the majority of the observation. This fact can be exploited to better model the transient behavior of the detector response and subtract it from the data.

Before entering into the details of the algorithm used, it is useful to establish some technical terms. The PACS array is composed by $5\times 5$ spatial pixels which are called spaxels. For each one of these spaxels, an instantaneous spectrum is measured using 16 spectral pixels which are called spexels. A large wavelength range is then obtained by moving the grating to cover a different instantaneous spectrum for each grating position.

The pipeline described in \cite{Fadda2016} is optimized for observations with a single band. In cases like M~104 where two bands are observed consecutively, the fit of the transient along the entire observation can be done only after normalizing the science blocks to their respective telescope background and correcting for the transients due to the change of spectral range.

We therefore modified the sequence of the modules performed in the script given with HIPE~15. After correcting all the science blocks for dark, response, and flats, we output them and correct them externally using Python scripts. We first compute the transient due to the change of band by considering the last off-target position block of each spaxel. To improve the statistics, we include a small selection of blocks from the end of the observation where the signal appear stabilized  (see left panel of Fig.~\ref{fig:slice}) since many blocks are damaged by cosmic ray transients.
After normalizing each science block, we compute a first estimate of the long-term transient affecting the entire observation.

This first estimate is obtained for each spaxel by combining all the 16 spexels (see Fig.~\ref{fig:scan}). We use it to correct the signal and obtain a second estimate of the science block transient by coadding all the science blocks (after rejecting the ones affected by cosmic-induced transients). This produces models of science block transients for each spexel, as shown in the right side of Fig.~\ref{fig:slice}.

At this point, this correction is applied to each single spexel of the original signal and the long-term transient is obtained for each individual spexel (Fig.~\ref{fig:pixel}, top panel). By correcting the spexels for this transient (middle panel of Fig.~\ref{fig:pixel}), the effects of cosmic rays on the response of the detector become evident. If not corrected they can introduce a significant amount of noise in the final spectral cube. Moreover, since they also affect the signal in the off-target positions, the subtraction of the off-target signal can lead to further inaccuracy in the final result.

The cosmic induced transients are corrected using the module {\it specTransCorr} in the transient correction pipeline of HIPE~15 as explained in \citet{Fadda2016}. This module is very efficient in removing the transients (see bottom panel of Fig.~\ref{fig:pixel}). Transients are identified by finding discontinuities in the flux and fitting the transient behaviour of the response after the discontinuity. The few readouts at the beginning of the transients are masked by the pipeline.

The data are finally reprojected into a spectral cube with a spatial pixel of 3~arcseconds. During this last step, a further rejection of outliers is performed by the pipeline using the redundancy of data in the same sky position.

The described process is not only able to eliminate most of the gradient visible in the archival data, but it also allows one to reconcile the two scans over the observed region. The usage of the telescope background as absolute flux calibrator is also very important. As visible in the ratio of signal over telescope background, the signal calibrated with the calibration blocks (the technique used in the archival product) grossly overestimates the flux (by approximately 55\%, in this case). We remind that the flux calibration using the telescope background is the standard process used for the chop/nod observations and it has been verified with sky flux calibrators. Also, as shown in \citet{Sutter2022}, the recalibration of unchopped data using the telescope background as absolute reference agrees very well with measurements done with SOFIA/FIFI-LS in NGC~7331~\citep[SOFIA/FIFI-LS uses planets, moons, and asteroids as absolute flux calibrators, see][]{Fadda2019}.

\section{Measurements}
\label{sec:measurements}
This appendix presents the measured fluxes in the \CII, CO, and optical lines, as well as the quantities derived from the SED fitting used in the figures of the paper (see Table~\ref{tab:sample}). For measurements determined using the CIGALE SED models (PAH 7.7\micron\ and 11.3\micron, FIR, and UV$_{\rm{ att}}$) we assume 10\% errors due to the uncertainties introduced during the modelling process.

\bibliography{main}{}
\bibliographystyle{aasjournal}

% Table with measurements
\movetabledown=4.7cm
\begin{rotatetable}
\begin{deluxetable*}{lccccccccccccccc}
\label{tab:sample}
\tabcolsep=0.1cm
\tablewidth{0pt}
\tabletypesize{\scriptsize}
\tablecaption{M104 aperture measurements}
\tablehead{
\colhead{Aper-} &
\colhead{Coordinates} &
\colhead{[CII]$_{158\mu m}$} &
\colhead{$^{12}$CO$_{J=0\rightarrow 1}$}&
\colhead{$\nu F_{\nu}$ ratio}&
\multicolumn{2}{c}{PAH}&
\colhead{FIR}&
\colhead{H$\beta$}&
\colhead{[OIII]$_{5008A}$}&
\colhead{H$\alpha$}&
\colhead{[NII]$_{6585A}$}&
\colhead{[SII]$_{6718,32A}$}&
\colhead{EW(H$\alpha$)}&
\colhead{UV att}\\
ture&
&
\colhead{W/m$^{2}$} &
\colhead{W/m$^{2}$} &
\colhead{70$\mu$m/100$\mu$m}&
7.7$\mu$m&11.3$\mu$m&8-1000$\mu$m&
\colhead{erg/s/cm$^2$}&
\colhead{erg/s/cm$^2$}&
\colhead{erg/s/cm$^2$}&
\colhead{erg/s/cm$^2$}&
\colhead{erg/s/cm$^2$}&
\colhead{\AA}&
\colhead{erg/s/kpc$^2$}\\
\colhead{ID} &
\colhead{J2000} &
\colhead{$\times10^{-21}$}&
\colhead{$\times10^{-18}$}&
\colhead{}&
\colhead{10$^6L_\odot$}&
\colhead{10$^6L_\odot$}&
\colhead{10$^6L_\odot$}&
\colhead{$\times10^{-16}$}&
\colhead{$\times10^{-16}$}&
\colhead{$\times10^{-16}$}&
\colhead{$\times10^{-16}$}&
\colhead{$\times10^{-16}$}&
&\colhead{$\times10^{40}$}
}
\startdata
N1&12:40:09.04 -11:37:20.4&23.2$\pm$0.5&18.3$\pm$0.7&0.41$\pm$0.06&2.47&1.21&37.1&2.6$\pm$0.8&2.4$\pm$0.6&18.3$\pm$0.9&10.4$\pm$0.7&9.5$\pm$1.3&7.3$\pm$0.4&4.65\\
N2&12:40:08.04 -11:37:14.1&18.7$\pm$0.7&12.4$\pm$0.4&0.42$\pm$0.07&2.01&1.13&31.1&0.9$\pm$0.9&2.7$\pm$0.6&8.7$\pm$0.8&7.7$\pm$0.6&7.4$\pm$0.9&2.8$\pm$0.3&1.76\\
N3&12:40:07.04 -11:37:11.5&13.6$\pm$0.7&10.4$\pm$0.4&0.44$\pm$0.08&1.50&0.83&24.1&0.9$\pm$0.8&1.9$\pm$0.6&8.4$\pm$0.7&5.5$\pm$0.6&4.8$\pm$0.5&2.5$\pm$0.2&1.88\\
N4&12:40:06.26 -11:37:10.4&14.1$\pm$0.0&10.8$\pm$0.4&0.44$\pm$0.09&1.29&0.74&19.9&  &2.0$\pm$0.7&5.0$\pm$0.6&4.7$\pm$0.5&4.3$\pm$8.2&1.2$\pm$0.2&1.79\\
N5&12:40:05.42 -11:37:10.2&21.2$\pm$0.4&10.1$\pm$0.7&0.44$\pm$0.10&1.14&0.71&16.9&1.9$\pm$1.2&2.0$\pm$0.7&13.2$\pm$1.1&7.0$\pm$0.8&6.4$\pm$0.7&2.8$\pm$0.2&1.23\\
N6&12:40:04.73 -11:37:09.8&9.9$\pm$1.0&11.8$\pm$0.4&0.45$\pm$0.11&1.11&0.70&16.9& & &7.2$\pm$1.5&6.8$\pm$1.0&6.7$\pm$1.1&1.3$\pm$0.3&0.38\\
N7&12:40:03.96 -11:37:08.8&8.4$\pm$3.9&9.7$\pm$0.3&0.44$\pm$0.11&1.07&0.71&16.6& & &8.6$\pm$1.1&5.8$\pm$0.8&4.7$\pm$0.7&1.3$\pm$0.2&0.39\\
N8&12:40:03.13 -11:37:09.1&7.1$\pm$1.0&10.5$\pm$0.3&0.47$\pm$0.11&1.15&0.84&18.4& &1.5$\pm$1.1&6.9$\pm$1.1&6.4$\pm$0.9&5.1$\pm$0.7&0.7$\pm$0.1&0.49\\
N9&12:40:02.27 -11:37:08.5&9.6$\pm$0.6&8.8$\pm$0.3&0.44$\pm$0.11&1.06&0.90&17.3&  & &4.1$\pm$1.0&7.1$\pm$0.8&5.7$\pm$0.9&0.3$\pm$0.1&0.41\\
N10&12:40:01.04 -11:37:09.1&13.8$\pm$1.6&10.3$\pm$0.3&0.48$\pm$0.11&1.16&1.06&18.3& &2.0$\pm$1.5&7.3$\pm$2.0&8.9$\pm$1.6&7.2$\pm$1.7&0.4$\pm$0.1&0.68\\
N11&12:40:00.17 -11:37:08.8&27.5$\pm$1.3&10.7$\pm$0.3&0.49$\pm$0.09&1.43&1.50&21.8& & &15.5$\pm$1.8&12.2$\pm$1.5&9.3$\pm$1.9&0.6$\pm$0.1&0.62\\
N12&12:39:58.79 -11:37:08.8&15.7$\pm$1.4&8.0$\pm$0.3&0.52$\pm$0.10&1.42&1.61&22.6&  &  &6.5$\pm$1.2&7.7$\pm$1.1&5.0$\pm$1.6&0.3$\pm$0.1&0.63\\
N13&12:39:57.74 -11:37:09.2&23.8$\pm$1.3&9.3$\pm$0.3&0.46$\pm$0.09&1.28&1.13&21.9& & &13.0$\pm$2.0&11.3$\pm$1.5&9.1$\pm$0.9&0.9$\pm$0.1&0.60\\
N14&12:39:56.81 -11:37:10.4&3.2$\pm$0.5&8.8$\pm$0.4&0.47$\pm$0.11&1.06&0.96&17.6&  & &2.2$\pm$1.8&7.4$\pm$1.7&5.0$\pm$0.8&0.2$\pm$0.1&0.47\\
N15&12:39:55.94 -11:37:11.5&4.7$\pm$1.2&8.5$\pm$0.4&0.42$\pm$0.10&1.07&0.83&18.1&  & &7.0$\pm$1.6&9.3$\pm$1.1&5.7$\pm$0.8&0.7$\pm$0.2&0.56\\
N16&12:39:55.01 -11:37:11.1&5.1$\pm$0.1&9.3$\pm$0.3&0.44$\pm$0.10&1.14&0.79&18.1&  & &5.4$\pm$1.6&9.0$\pm$1.4&6.8$\pm$1.6&0.7$\pm$0.2&0.46\\
N17&12:39:54.07 -11:37:12.5&16.9$\pm$1.5&10.0$\pm$0.4&0.38$\pm$0.09&1.28&0.80&20.6&  & &4.8$\pm$1.1&5.9$\pm$1.0&5.4$\pm$0.9&0.9$\pm$0.2&0.17\\
N18&12:39:53.27 -11:37:13.4&16.0$\pm$1.2&10.7$\pm$0.4&0.37$\pm$0.09&1.37&0.82&21.6&  & &2.8$\pm$0.7&5.3$\pm$0.7&4.7$\pm$0.5&0.6$\pm$0.2&2.28\\
N19&12:39:52.41 -11:37:14.8&25.2$\pm$1.1&10.6$\pm$0.3&0.39$\pm$0.08&1.48&0.86&24.0& & &9.8$\pm$0.7&8.2$\pm$0.6&6.7$\pm$0.5&2.4$\pm$0.2&1.91\\
N20&12:39:51.38 -11:37:17.4&20.1$\pm$1.0&11.0$\pm$0.4&0.42$\pm$0.07&1.91&1.10&30.2&1.3$\pm$0.8&1.6$\pm$0.7&10.2$\pm$0.8&8.4$\pm$0.6&7.2$\pm$0.5&2.9$\pm$0.2&1.91\\
N21&12:39:50.62 -11:37:18.0&17.9$\pm$0.6&9.7$\pm$0.4&0.42$\pm$0.06&2.09&1.18&33.0&1.4$\pm$0.8&2.5$\pm$0.7&10.4$\pm$0.8&8.3$\pm$0.6&8.2$\pm$1.7&3.5$\pm$0.3&1.71\\
N22&12:39:50.10 -11:37:25.9&25.0$\pm$8.9&16.5$\pm$0.5&0.41$\pm$0.06&2.70&1.29&31.2&2.1$\pm$1.2&2.0$\pm$0.5&13.5$\pm$0.8&8.6$\pm$0.6&8.3$\pm$0.4&5.7$\pm$0.4&5.05\\
N23&12:39:49.28 -11:37:24.4&20.1$\pm$0.5&15.1$\pm$0.0&0.37$\pm$0.07&2.12&1.03&25.1&  &  &  &  &  &  &1.29\\
L1&12:40:00.82 -11:37:22.8&14.0$\pm$0.3&  &0.69$\pm$0.13&0.91&1.43&17.5&  & &15.3$\pm$9.7&28.4$\pm$7.2&16.2$\pm$5.2&0.3$\pm$0.2&0.34\\
L2&12:40:02.01 -11:37:23.4&14.9$\pm$0.9&  &0.66$\pm$0.11&0.82&1.05&17.0&  &7.6$\pm$5.7&20.6$\pm$10.6&32.0$\pm$5.0&17.3$\pm$2.3&0.6$\pm$0.3&0.30\\
L3&12:40:03.03 -11:37:22.8&8.7$\pm$0.1&  &0.63$\pm$0.12&0.67&0.86&14.2&  &9.3$\pm$6.5&18.5$\pm$10.9&27.3$\pm$3.9&15.9$\pm$2.7&0.9$\pm$0.5&0.33\\
L4&12:40:04.08 -11:37:22.7&12.4$\pm$0.6&  &0.58$\pm$0.13&0.62&0.76&13.3&  &7.9$\pm$3.3&9.5$\pm$3.3&17.1$\pm$2.5&12.6$\pm$1.8&0.7$\pm$0.2&0.20\\
L5&12:40:05.15 -11:37:22.4&14.5$\pm$0.8&  &0.51$\pm$0.13&0.63&0.72&11.8&  & &3.6$\pm$2.4&7.2$\pm$1.7&4.9$\pm$1.2&0.4$\pm$0.3&0.26\\
L6&12:40:06.08 -11:37:22.0&8.4$\pm$0.7&  &0.48$\pm$0.11&0.82&0.79&14.3&  &4.7$\pm$1.2&4.6$\pm$1.3&7.8$\pm$1.0&5.4$\pm$0.6&0.8$\pm$0.2&0.18\\
L7&12:40:06.95 -11:37:22.1&12.6$\pm$1.1&  &0.45$\pm$0.08&1.24&1.00&17.7& &4.0$\pm$2.3&6.3$\pm$1.3&11.5$\pm$1.0&9.2$\pm$0.6&1.2$\pm$0.3&0.22
\enddata
\end{deluxetable*}
\end{rotatetable}

\movetabledown=5.2cm
\begin{rotatetable}
\begin{deluxetable*}{lcccccccccccccc}
\label{tab:sample2}
\tablenum{B1}
\tabcolsep=0.1cm
\tablewidth{0pt}
\tabletypesize{\scriptsize}
\tablecaption{(continued)}
\tablehead{
\colhead{Aper-} &
\colhead{Coordinates} &
\colhead{[CII]$_{158\mu m}$} &
\colhead{$^{12}$CO$_{J=0\rightarrow1}$}&
\colhead{$\nu F_{\nu}$ ratio}&
\multicolumn{2}{c}{PAH}&
\colhead{FIR}&
\colhead{H$\beta$}&
\colhead{[OIII]$_{5008A}$}&
\colhead{H$\alpha$}&
\colhead{[NII]$_{6585A}$}&
\colhead{[SII]$_{6718,32A}$}&
\colhead{EW(H$\alpha$)}&
\colhead{UV att}\\
ture&
&
\colhead{W/m$^{2}$} &
\colhead{W/m$^{2}$} &
\colhead{70$\mu$m/100$\mu$m}&
7.7$\mu$m&11.3$\mu$m&8-1000$\mu$m&
\colhead{erg/s/cm$^2$}&
\colhead{erg/s/cm$^2$}&
\colhead{erg/s/cm$^2$}&
\colhead{erg/s/cm$^2$}&
\colhead{erg/s/cm$^2$}&
\colhead{\AA}&
\colhead{erg/s/kpc$^2$}\\
\colhead{ID} &
\colhead{J2000} &
\colhead{$\times10^{-21}$}&
\colhead{$\times10^{-18}$}&
\colhead{}&
\colhead{10$^6L_\odot$}&
\colhead{10$^6L_\odot$}&
\colhead{10$^6L_\odot$}&
\colhead{$\times10^{-16}$}&
\colhead{$\times10^{-16}$}&
\colhead{$\times10^{-16}$}&
\colhead{$\times10^{-16}$}&
\colhead{$\times10^{-16}$}&
&\colhead{$\times10^{40}$}
}
\startdata
S1&12:40:08.10 -11:37:26.4&20.0$\pm$0.2&10.3$\pm$0.4&0.41$\pm$0.07&2.16&1.11&26.4& &2.0$\pm$0.4&7.5$\pm$0.6&7.3$\pm$0.5&6.3$\pm$0.5&3.5$\pm$0.3&4.50\\
S2&12:40:07.11 -11:37:31.0&10.7$\pm$0.0&7.5$\pm$0.4&0.43$\pm$0.09&1.42&0.79&17.9&1.7$\pm$0.4&1.3$\pm$0.4&12.6$\pm$0.5&7.2$\pm$0.4&6.4$\pm$0.3&8.2$\pm$0.4&0.36\\
S3&12:40:06.19 -11:37:33.0&9.6$\pm$1.2&9.4$\pm$0.3&0.44$\pm$0.10&1.25&0.67&16.5& &0.9$\pm$0.4&3.6$\pm$0.4&4.1$\pm$0.4&4.2$\pm$0.3&2.7$\pm$0.3&0.25\\
S4&12:40:05.22 -11:37:34.6&12.9$\pm$0.5&8.6$\pm$0.3&0.42$\pm$0.11&1.22&0.64&15.5&1.1$\pm$0.4&1.0$\pm$0.2&8.5$\pm$0.8&5.6$\pm$0.9&5.1$\pm$0.6&35.4$\pm$5.1&0.27\\
S5&12:40:04.27 -11:37:35.0&8.4$\pm$0.7&8.1$\pm$0.3&0.45$\pm$0.11&1.19&0.69&15.7&  &2.0$\pm$0.8&4.1$\pm$0.9&5.7$\pm$1.1&5.1$\pm$0.6&4.1$\pm$1.0&0.28\\
S6&12:40:03.10 -11:37:35.3&3.6$\pm$0.8&10.0$\pm$0.3&0.48$\pm$0.10&1.29&0.78&17.3&  &1.4$\pm$0.7&2.3$\pm$1.0&5.1$\pm$0.9&4.1$\pm$0.6&1.0$\pm$0.5&0.36\\
S7&12:40:01.95 -11:37:36.3&5.5$\pm$0.8&9.5$\pm$0.3&0.48$\pm$0.11&1.43&0.91&17.9& &1.5$\pm$0.6&4.8$\pm$0.6&6.9$\pm$0.6&5.4$\pm$0.6&1.3$\pm$0.2&0.27\\
S8&12:40:00.89 -11:37:36.6&13.0$\pm$1.0&8.3$\pm$0.5&0.46$\pm$0.10&1.76&1.14&20.1& &1.3$\pm$0.8&6.1$\pm$1.1&6.8$\pm$0.9&6.1$\pm$0.8&2.3$\pm$0.4&0.39\\
S9&12:39:59.70 -11:37:37.3&6.9$\pm$0.5&6.6$\pm$0.3&0.53$\pm$0.11&2.21&1.72&22.6& &1.9$\pm$0.7&4.7$\pm$0.8&5.5$\pm$0.6&4.7$\pm$0.6&0.7$\pm$0.1&0.61\\
S10&12:39:58.78 -11:37:37.3&11.6$\pm$0.3&8.6$\pm$0.5&0.52$\pm$0.11&2.00&1.52&20.3&  &3.0$\pm$1.4&4.7$\pm$1.1&5.5$\pm$0.8&4.5$\pm$0.6&0.8$\pm$0.2&0.55\\
S11&12:39:57.88 -11:37:38.3&6.6$\pm$1.1&8.1$\pm$0.3&0.39$\pm$0.12&1.55&1.06&17.7& &1.2$\pm$0.4&2.6$\pm$0.4&4.8$\pm$1.0&3.7$\pm$0.7&2.8$\pm$0.5&0.34\\
S12&12:39:57.09 -11:37:37.3&6.7$\pm$0.4&8.6$\pm$0.3&0.42$\pm$0.11&1.41&0.93&17.0&1.5$\pm$0.7&1.4$\pm$0.4&8.8$\pm$0.7&6.8$\pm$0.6&6.1$\pm$0.6&2.3$\pm$0.2&0.29\\
S13&12:39:56.21 -11:37:36.6&6.1$\pm$0.5&11.6$\pm$0.3&0.41$\pm$0.11&1.28&0.81&16.3& &1.9$\pm$0.4&3.9$\pm$0.6&5.6$\pm$0.5&4.8$\pm$0.6&1.3$\pm$0.2&0.24\\
S14&12:39:55.31 -11:37:36.0&15.6$\pm$1.2&12.7$\pm$0.8&0.41$\pm$0.11&1.23&0.74&16.8&0.8$\pm$0.7&1.8$\pm$0.6&5.9$\pm$0.7&7.0$\pm$0.6&6.2$\pm$0.4&2.1$\pm$0.2&0.35\\
S15&12:39:54.36 -11:37:36.3&8.7$\pm$0.5&11.5$\pm$0.3&0.39$\pm$0.11&1.25&0.69&15.9&1.1$\pm$0.8&1.4$\pm$0.5&5.7$\pm$0.5&6.2$\pm$0.5&5.8$\pm$0.6&6.1$\pm$0.6&0.29\\
S16&12:39:53.35 -11:37:35.3&14.2$\pm$1.1&16.2$\pm$0.5&0.34$\pm$0.10&1.30&0.74&16.9& &0.9$\pm$0.6&2.6$\pm$0.5&4.1$\pm$0.4&3.5$\pm$0.2&1.5$\pm$0.3&0.24\\
S17&12:39:52.36 -11:37:34.6&20.5$\pm$1.0&14.4$\pm$0.4&0.41$\pm$0.09&1.53&0.77&20.8&1.4$\pm$0.5&0.7$\pm$0.3&10.8$\pm$0.5&6.1$\pm$0.5&5.3$\pm$0.3&7.6$\pm$0.4&0.25\\
S18&12:39:51.55 -11:37:32.0&12.4$\pm$0.4&12.2$\pm$0.4&0.40$\pm$0.07&1.89&0.99&22.4& &1.1$\pm$0.4&5.9$\pm$0.6&5.5$\pm$0.5&5.3$\pm$0.4&3.4$\pm$0.4&0.47\\
S19&12:39:50.76 -11:37:29.3&21.1$\pm$0.9&12.0$\pm$0.4&0.42$\pm$0.06&2.54&1.22&30.6&1.2$\pm$0.7&1.8$\pm$0.5&9.8$\pm$0.7&6.4$\pm$0.5&6.4$\pm$0.5&4.8$\pm$0.4&5.06\\
S20&12:39:47.78 -11:37:25.6&4.5$\pm$0.5&5.0$\pm$0.4&0.30$\pm$0.15&0.84&0.51&13.3&  &  &  &  &  &  &3.14\\
S21&12:39:46.82 -11:37:25.0&5.5$\pm$0.8&4.3$\pm$0.3&0.27$\pm$0.21&0.62&0.35&9.5&  &  &  &  &  &  &2.10\\
S22&12:40:10.73 -11:37:20.5&18.9$\pm$1.6&7.4$\pm$0.2&0.34$\pm$0.11&1.24&0.66&18.5&  &  &  &  &  &  &4.21\\
Nuc&12:39:59.41 -11:37:22.0&31.4$\pm$2.4&2.1$\pm$1.5&1.10$\pm$0.10&3.11&2.85&53.6& & &110.6$\pm$17.5&695.0$\pm$48.0&302.6$\pm$38.5&0.4$\pm$0.1&0.65\\
R1&12:39:58.61 -11:37:21.8&14.9$\pm$1.4&1.4$\pm$0.2&0.83$\pm$0.11&1.35&2.25&25.8& &31.8$\pm$6.8&38.6$\pm$15.3&78.1$\pm$10.7&36.8$\pm$5.6&0.5$\pm$0.2&0.31\\
R2&12:39:57.34 -11:37:23.7&8.9$\pm$0.9&  &0.65$\pm$0.11&0.85&1.16&18.1&  & &17.6$\pm$7.8&45.9$\pm$7.8&22.0$\pm$3.5&0.5$\pm$0.2&0.36\\
R3&12:39:56.29 -11:37:24.1&11.4$\pm$0.8&  &0.58$\pm$0.15&0.65&0.90&13.0&  &11.3$\pm$3.8&4.8$\pm$4.8&17.4$\pm$4.2&8.5$\pm$2.2&0.2$\pm$0.2&0.26\\
R4&12:39:55.20 -11:37:24.0&8.9$\pm$0.8&  &0.56$\pm$0.13&0.61&0.81&12.6& &10.6$\pm$6.1&14.8$\pm$3.2&22.6$\pm$2.7&15.1$\pm$2.6&1.0$\pm$0.2&0.26\\
R5&12:39:54.36 -11:37:24.4&9.2$\pm$1.1&  &0.47$\pm$0.14&0.62&0.78&11.4&  &6.9$\pm$4.3&5.4$\pm$2.8&10.9$\pm$1.9&6.4$\pm$1.2&0.5$\pm$0.3&0.18\\
R6&12:39:53.44 -11:37:24.4&18.6$\pm$1.4&  &0.41$\pm$0.13&0.77&0.85&12.4&  & &1.3$\pm$2.0&5.9$\pm$1.3&3.7$\pm$2.2&0.2$\pm$0.3&0.24\\
R7&12:39:52.58 -11:37:24.8&11.8$\pm$0.7&  &0.43$\pm$0.10&1.05&0.94&16.9&  &3.2$\pm$2.1&5.8$\pm$1.2&10.5$\pm$1.0&7.6$\pm$0.8&1.1$\pm$0.2&0.31
\enddata
\end{deluxetable*}
\end{rotatetable}



\end{document}